%% file: main.tex
\documentclass[fleqn,usenatbib]{rasti}
\usepackage{fix-cm}

\usepackage{newtxtext,newtxmath}
\usepackage{txfonts}

\usepackage[T1]{fontenc}

\DeclareRobustCommand{\VAN}[3]{#2}
\let\VANthebibliography\thebibliography
\def\thebibliography{\DeclareRobustCommand{\VAN}[3]{##3}\VANthebibliography}

\usepackage{graphicx}	
\usepackage{amsmath}	
\usepackage{xcolor}	    
\usepackage{cleveref}   
\usepackage{array}

\providecommand{\code}[1]{\texttt{\detokenize{#1}}}
\providecommand{\prompt}[1]{\texttt{\detokenize{#1}}}

\providecommand{\orcit}[1]{\protect\href{https://orcid.org/#1}{\protect\includegraphics[width=8pt]{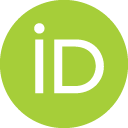}}}

\newcommand\itemc{\item[\textbf{$\square$}]}

\newcommand\itemr{\item[\textbf{$\medcirc$}]}

\title[The Role of LLMs in Astronomy Research]{What is the Role of Large Language Models in the Evolution of Astronomy Research?}

\input{authors} 

\date{Accepted XXX. Received YYY; in original form ZZZ}

\pubyear{2024}

\begin{document}
\label{firstpage}
\pagerange{\pageref{firstpage}--\pageref{lastpage}}
\maketitle

\begin{abstract}
ChatGPT and other state-of-the-art large language models (LLMs) are rapidly transforming multiple fields, offering powerful tools for a wide range of applications. These models, commonly trained on vast datasets, exhibit human-like text generation capabilities, making them useful for research tasks such as ideation, literature review, coding, drafting, and outreach. We conducted a study involving 13 astronomers at different career stages and research fields to explore LLM applications across diverse tasks over several months and to evaluate their performance in research-related activities. This work was accompanied by an anonymous survey assessing participants' experiences and attitudes towards LLMs. We provide a detailed analysis of the tasks attempted and the survey answers, along with specific output examples. Our findings highlight both the potential and limitations of LLMs in supporting research while also addressing general and research-specific ethical considerations. We conclude with a series of recommendations, emphasizing the need for researchers to complement LLMs with critical thinking and domain expertise, ensuring these tools serve as aids rather than substitutes for rigorous scientific inquiry.
\end{abstract}





\section{Introduction}

Language models are probabilistic representations of the natural human language. The first such models appeared in the 1980s and were purely statistical, based on the frequencies of co-occurrence of phrases of different lengths, so-called {\it n-grams}. The power of language models has grown dramatically in the last 5 years with the development of transformers \citep{Transformers}.
A transformer is an advanced machine learning model that improves how data sequences are analyzed and generated by simultaneously considering the full context of the input and focusing on the most relevant parts, enhancing performance in language-related tasks. This new architecture, combined with neural networks and large datasets (frequently scraped from the internet), has led to the development of the current (as of mid-2024) generation of Large Language Models (LLMs) as advanced artificial intelligence (AI) systems capable of parsing and generating human-like text. 

Current LLMs represent a tremendous technological leap. Unlike other leaps of technology, which were expensive and very few people had access to them at first, LLMs are (in many cases) free and available to (almost) anyone with an internet connection. As a result, their adoption has also been incredibly fast - ChatGPT reached 100 million users within only two months after its launch \citep{hu2023} while other services of similar popularity took between 9 months and 4.5 years to reach the same userbase\footnote{https://www.visualcapitalist.com/threads-100-million-users/}. Unlike other technologies primarily automating repetitive work, LLMs excel at creative work like writing, coding, and generating ideas. Finally, the literature surrounding productivity boosts attributed to large language models (LLMs) suggests that there are significant improvements in various tasks, with estimates ranging from 20\% to 80\% in productivity boosts across different sectors \citep[][]{developerproductivity, productivity1, productivity2} compared to the 18-22\% improvement brought on by steam power \citep{ATACK2008185}.

Also, unlike other technological advances, LLMs behave very differently from traditional technology. As discussed in \citep[][p. 65-67]{Mollick2024}, LLMs defy our expectations for software functionality: software should produce predictable, reliable, logical outcomes, and LLMs do not. On the contrary, LLMs (at the moment, at least) struggle with tasks that software is generally good at, such as repeating tasks, reproducing facts, and performing calculations. In contrast, they perform well on tasks we consider uniquely human: writing, chatting, analyzing, coding, brainstorming, and creating. In addition, LLMs have "features" that we do not encounter in other technologies: they confidently invent facts and make mistakes, aka hallucinate. These latter features make many researchers suspicious of using them in a research context. 

LLMs are expected to have a large impact on a wide range of professions, especially those in highly paid and creative fields, including STEM. A 2023 McKinsey report \citep{Ellingrud2023} predicts that 16\% of the hours worked by STEM professionals today will be automated by generative AI by 2030. 
Undoubtedly, a growing number of astronomers have tried LLMs, and many frequently utilize them for a wide range of tasks, sometimes secretly and sometimes openly. Several universities have communicated recommendations to staff and students on the appropriate use of LLMs in teaching. Still, only a handful of professional astronomical entities and journals have published guidance on the recommended use of LLMs specifically focused on research manuscripts and the peer-review process, leaving a vacuum for other professional tasks. \citet{astarita2024} have shown that the uptake of LLMs in journal articles is likely high. While many raise concerns about how LLMs may change education \citep[e.g.,][]{Fyfe2022, Perkins2023}, we are only starting to consider the implications on scientific research \citep[e.g.,][]{Chubb2021, Dwivedi2023}. 

Like any new technology, LLMs have caveats and deserve technical and ethical scrutiny. Some of the questions raised are: How accurate are their responses in the context of tasks done by researchers? Are they capable of writing code suitable for scientific tasks, and is the code accurate?  Does using LLMs for editorial corrections produce the voice expected in academic writing? Does the use of LLMs compromise scientific integrity? Where lies the ethical boundary of using LLMs? Is the quality of LLM responses correlated with the cost of the tools? How should we acknowledge the use of such tools? In light of this growing use of LLMs to conduct research, it appeared essential to explore tasks particularly well suited to LLMs, identify good practices, raise awareness of their potential impacts and limitations, and make recommendations on their responsible use.

To explore the potential uses of LLMs in a research context, we initiated a study involving a group of research astronomers at different career stages. The group included Master's and PhD students, post-docs, and staff scientists. The research areas and the prior experience level also varied across the group. The purpose of this study was to investigate as users the following four main areas: (a) the potential applications and limitations in using LLMs to do astronomy research and daily tasks, (b) the positive and negative effects of LLMs on the science research environment, (c) the necessary skills to effectively utilize LLMs and (d) the ethical considerations associated with the use of LLMs in science. We stress that we are not researching the technology but considering LLMs as a technology to assist us in our research.

This paper is organized as follows. Sect.~\ref{sec:exp-approach} describe our study approach, the major tasks we considered part of an astronomer's work, and the services we explored to tackle these tasks. To extend our viewpoints, we collected our colleagues' input in a survey we describe in Sect.~\ref{sec:survey-results} (and Appendix \ref{sec:survey_questions}). Sect.~\ref{sec:use-cases} describes the experiments and results for the individually identified use cases. Following the presentation of our findings, Sect.~\ref{sec:discussion} summarizes the main benefits and limitations, discusses legal and ethical concerns, and Sect.\ref{sec:recommendations} draws general tips for using LLMs for research before we summarize our study in Sect.~\ref{sec:summary}.

\section{Experimental approach}\label{sec:exp-approach}

In October 2023, we called researchers within the Max Planck Institute of Astronomy to join a group to explore LLMs' potential, limitations, and impacts on scientific work. The call was open to institute members from all career levels and scientific departments. Approximately 20 people volunteered, $\sim$15 of whom joined the work of the group in some capacity or another. These participants were primarily graduate students and postdocs but also staff with diverse data science expertise and various scientific interests. Only a minority of this group was familiar with the technical aspects of LLMs; however, delving into the internal details of LLMs was not the group's task. The work of the group consisted of a series of regular meetings chaired by the lead author, where the group discussed logistics, ideas, and results from various experiments.

The group identified a set of LLM services available at the time (November 2023), and participants volunteered to try one or more of them. Specifically, the services identified and explored are listed in \cref{tab:services_used} with a more detailed technical description provided in Appendix \ref{sec:services}. This list is not exhaustive and reflects LLMs' capabilities at the time of writing.

\begin{table}
    \centering
    \caption{Services we used during this study. Detailed characteristics in Appendix \ref{sec:services}}
    \label{tab:services_used}
    \begin{tabular}{lll}
        \hline
        Which                   & Based upon$^\star$          & Short description         \\
        \hline
        {\code{ChatGPT}}        & GPT-4/Plus                  & Chatbot                   \\
        {\code{Perplexity AI}}  & pplx-Sonar, GPT-4, Claude 3 & Chatbot                   \\
        {\code{Bard/Gemini}}    & PaLM 2/Gemini               & Chatbot                   \\
        {\code{You.com}}        & GPT-4, Claude 3, Gemini, Llama,  & Chatbot                   \\
                                & Mistral, and proprietary models       &                           \\
        {\code{GitHub Copilot}} & Codex + GPT-3.5             & Coding assistant          \\
        {\code{Grammarly}}      & Proprietary model                  & Writing assistant         \\
        {\code{DeepL/Write}}    & Proprietary model                   & Writing assistant         \\
        {\code{Codeium}}        & GPT-3/GPT4                 & Coding assistant          \\
        {\code{Midjourney}}     & Stable diffusion            & Image generation          \\
        {\code{Dall-E}}         & Stable diffusion            & Image generation          \\
        \hline
    \end{tabular}
    $^\star$ Models available at the time we did our study. Some services may have changed since.
\end{table}

As one can see from the list in \cref{tab:services_used}, the range of services is broad. It includes conversational interfaces such as ChatGPT, models trained for specific tasks (e.g., Grammarly, Copilot), and LLM-powered search engines (e.g., Perplexity, you.com). We also need to highlight that this is not, by any means, an exhaustive list of applications currently on the market. Neither is this an endorsement of these services. In each category above, there is solid competition, and we encourage readers to explore alternatives to find one that fits their budget and work style. We also tested image generation models, Midjourney and DALL-E.

Several services we used offer both a free and a paid use tier. Our group received a budget of $\sim$1000 Euro for this project, and we acquired licenses for ChatGPT, Perplexity, and Grammarly for the duration of the exploratory work. Most accounts were open between 3 and 5 months. Additionally, GitHub provides students and academics free access to Copilot for educational purposes. Participants who used Copilot were able to access it through this program. Finally, our institute provides staff with access to DeepL Pro. We only used the free tier services for other providers.

The first phase of the experiment was ideation. Participants who had used LLMs in research tasks shared experiences, and the group brainstormed other potential opportunities. Participants were encouraged to ``invite LLMs to every conversation'', i.e., to try to incorporate LLMs in all tasks to learn about their strengths, weaknesses, and unique features. In this way, we attempted to map the "Jagged Frontier of AI" for astronomy researchers, as \citet{DellAcqua2023} have named the line between tasks that LLMs are good at and those they fail at. We identified primary use cases that we grouped into the following six categories:

\begin{itemize}
    \item{{\bf Text generation and improvement} including tasks such as drafting emails, letters, papers, and proposals; editing and proofreading academic manuscripts; formatting citations and bibliographies; content generation for teaching, including quizzes, worksheets, presentations, and syllabuses (also see item on Teaching below); generating simple explanations of complex concepts; brainstorming topics and ideas for content. Tasks in this category also include professional correspondence.}

    \item{{\bf Summarizing and analyzing content} including tasks such as summarizing individual documents or multiple documents to extract key points; identifying gaps in analysis; identifying similarities and differences across publications; creating summaries of existing research and literature; summarizing papers for various audiences; identifying trends in the field. Most of these tasks are especially well suited to literature review. With access to the internet now, LLMs can identify what datasets specific papers used, look up citations, and summarize them. An extension of this process is creating tailored models for discussing literature and research \citep[e.g.][and related work by Universe TBD\footnote{https://universetbd.org/}]{Ciucă_2023, wu2024designingevaluationframeworklarge} however this is beyond the scope of our work.}

    \item{{\bf Software development} including tasks such as generating and analyzing code; debugging and refactoring code; code documentation and explanation; converting code between programming languages. With assistance from GitHub Copilot, \citet{developerproductivity} found developers were able to complete their tasks 55.8\% faster than without. LLMs can create pieces of scripts for complex tasks, and they can suggest ideas or approaches that might be outside the programmer’s direct scope/knowledge, which can lead to either a learning effect and/or improved code.}

    \item{{\bf Data Analysis and categorization} LLMs can process and analyze vast amounts of data, extracting key insights and information that would be challenging and time-consuming for humans to identify and aiding in the decision-making processes. Tasks here include processing and analyzing large datasets to extract insights, identifying trends and patterns in data, and categorizing and organizing data. General-purpose LLMs can assist in this task with some effort, but there is a growing number of offers for specialized assistants that focus on data exploration tasks (e.g., Deepnote).}

    \item{{\bf Teaching}: including tasks such as generating explanations of concepts for students of different levels, creating and refining materials through interactive prompts, creating educational content like quizzes and interactive learning materials, and using LLMs as interactive tools for learning new material. Significant debate exists in the educational community on using LLMs in teaching. As part of the educational system, we must provide students with basic AI literacy, even if we do not use LLMs in the educational process.}

    \item{{\bf Image and video generation}: Some large language models interconnect with image generation models, which can create images and videos from text descriptions. One can use these in illustrations, presentations, and outreach material. LLMs can also improve accessibility by creating text descriptions for images and videos. However, we must ensure such models do not replace or misuse the work of artists and illustrators.}
\end{itemize}

This manuscript focuses on the first four categories of applications and outreach because this is where the core expertise of our team lies. While we did some work on the applications of LLMs to teaching, we left such exploration to experts in these fields.

Following the ideation phase, group participants self-assigned themselves to test specific use cases. These tests were done over three months in the Spring of 2024. Regular meetings were held to discuss progress and coordinate the efforts. Participants documented their ongoing work in a shared document and wrote a description of their work, results, and general feedback at the end of the project (April 2024). These notes serve as the basis for the following sections. Appendix \ref{tab:services_used} contains excerpts from specific LLM interactions.

Since the group had a set of limited viewpoints and, perhaps, the biased attitude of early adopters, we also designed and circulated an anonymous survey to all institute members to probe the attitudes and uses of LLMs.

\section{Extending findings through a survey}\label{sec:survey-results}

A survey on the attitudes towards and uses of LLMs ran from March 11 to 28 among the employees of MPIA and collected 27 responses. The survey participants range from interns to professors. The questions and responses are detailed in Appendix \ref{sec:survey_questions}. 
This section discusses the parts of the survey that focused on general questions, ethics, and overall satisfaction. We will discuss the results of this survey in the relevant use-case sections throughout the next sections.  

Among the 27 survey participants, 74\% use LLMs at least several times a week. ChatGPT-3.5 and ChatGPT-4 are the most commonly used, with 89\% and 44\% of participants indicating that they have experience with them (multiple answers were possible). Following ChatGPT in the list of most frequently used assistants are Grammarly with 41\% and GitHub Copilot with 37\% (details in Appendix \ref{sec:survey_questions}). The most common tasks are coding (software development) with 92\% and writing assistance with 72\% of responses (multiple answers were possible). None of the participants find it difficult to use LLMs; on the contrary, 70\% of respondents find it easy or very easy to use them. 30\% of the respondents indicated some challenges in using LLMs. 

In terms of ethics, 78\% of participants said that credit should be given to LLMs in science publications, but 61\% say that they rarely do so at the moment. We discuss some possible reasons in Sect. \ref{sec:research-specific-concerns}. 

Overall, 48\% of participants were satisfied or very satisfied with the current capabilities of LLMs, 41\% were neutral about them, and 11\% were not or not at all satisfied. Satisfaction is the highest for software development: 77\% are satisfied or very satisfied with using LLMs in that context. Fewer participants are enthusiastic about their writing and problem-solving abilities: 50\% are satisfied or very satisfied, and 25\% are neutral. More than half of the respondents (59\%) indicate a desire to have access to the paid version of such tools, with ChatGPT and GitHub Copilot being the top choices of tools where paid access is desired (note that GitHub Copilot does not have general free access version but students and teachers can get educational access corresponding to an individual tier free of charge). Most participants (83\%) believe that LLMs will become an integral part of science, but only 42\% think they will be a good influence overall. 

The survey also asks the participants about their experience and satisfaction in specific use cases we detail below. We discuss those results in the relevant contexts.

\section{Use Cases of LLMs as a Tool for Research}\label{sec:use-cases}

The main focus of our experiments was to explore the use cases of LLMs in the research workflow of astronomers. Even though LLMs have multiple use cases in non-research tasks, this section focuses on tasks related to research activities. Specifically, we tested various models' abilities to generate and improve text, summarize text, write scientific code, and handle data analysis 
The specific experiments, their results, and conclusions based on these experiences are detailed below.

\subsection{Academic Writing}\label{subsec:writing}

Writing is essential for all academics and critical for our professional success. From professional communications to cover letters and research plans to manuscripts and funding proposals, written text is indispensable for connecting with colleagues, conveying our findings, obtaining jobs, and securing resources such as telescope/computational time and money. LLMs can assist researchers in various aspects of the writing process, such as generating text, improving readability, and providing suggestions for structure and content. LLM tools can also assist with language translation, grammar checking, and citation formatting.

From our internal survey (see Appendix \ref{sec:survey_questions}), we found that writing assistance is the second most common use case (72\% of responses) for LLMs (after coding assistance, 92\%). 59\% of the survey participants have used LLMs specifically for improving academic texts, but most estimate that the overall contribution of LLMs to their writing is small: only 23\% say that LLMs have contributed more than 20\% to their texts. However, there is a 0\% to 99\% spread in how users feel satisfied by the answers of LLMs. Such a result suggests a varying degree of accuracy in the answers, possibly reflecting the range of expectations, use cases, and prompting skills among the participants. 

Based on our tests, most current LLMs offer valuable assistance in understanding scientific concepts, software, and instruments. They can summarize internet searches on specific topics into bullet points, covering introductions, current research, challenges, and prospects. Different services vary in their abilities to cite peer-reviewed articles. LLM-enabled search engines do better. There is also a variation in the style of the response. For example, \texttt{Gemini} or \texttt{ChatGPT} rely on a mix of academic and popular science sources by default. Other platforms (e.g., You.com) allow users to specify a style explicitly. These tools often provide more readable and comprehensive information than Wikipedia articles, which are commonly the first resource for beginners. LLMs can also assist in drafting academic paper templates. This is particularly reliable for papers that employ well-established data sources and analysis techniques. This feature benefits students transitioning to academic writing, non-native English speakers, or anyone facing writer's block. However, our study participants found that the generated templates still require substantial elaboration in style and content before submission to a journal. Hence, the work of the student/researcher is far from being eliminated.

It remains important to be the "human in the loop," as there are always instances where responses from these tools require corrections. 
For example, when prompting GPT-3.5 with "In what situations can we use MHD? In what regimes does it break down?", it stated that "magneto-hydrodynamics assumes collisionless plasma" when the exact opposite is true. When informed about their errors, both chatbots immediately provided the correct output. Gemini also offers an option to cross-check responses with a Google search\footnote{Gemini cross-check with Google search is done after and does not help to construct the initial response}. When asking varied tools, "Who leads the Gaia DPAC?" Gemini answers Francois Mignard wrongly; GPT4 does not find answers and refers to the official Gaia/ESA pages; You.com (genius mode) replies correctly not only the name of the chair (and since when) but also the deputy chair correctly and cites sources.

It is difficult for a non-expert to spot these mistakes, as they are very technical. Therefore, if a certain piece of information is key to the user’s writing, it is essential to fact-check through a conventional search.

The clarity of our messages is critical in academic writing. Clear communication helps avoid misunderstandings and misinterpretations. LLMs can detect potentially unclear or ambiguous sentences and suggest rephrasing or simplifying these sentences to improve clarity. These models can also assess the overall structure and flow of the message and identify sections that needorganization or additional context. By ensuring coherence and cohesion, your message becomes more readable and impactful. LLMs can suggest alternative words or phrases to enrich your vocabulary. Overall, they help you express your ideas more precisely and effectively.

LLMs can also analyze a message's content to identify inappropriate or inconsistent tones and readability scores. They can flag sentences that may be too formal or informal. LLMs can also quickly draft a message that only needs some edits, filter incorrect tones, or highlight potentially unclear sentences. Interpersonal communications are complicated, especially in written forms; LLMs can aid by guaranteeing that our written messages convey the intended meaning, appropriate tone, and level of emotions. Such assistance lets us focus on the content and adjust the form/tone to ensure the message comes across clearly. Enhanced language adds depth and professionalism to your written communication. All these are critical in efficient written communications with colleagues. Such assistance can be beneficial for junior researchers who have had less exposure to formal academic communications. A drawback is that writing can become standardized or mechanical.

It is essential to acknowledge that English is not the first language of many -- if not most -- professional astronomers who may face bias and discrimination because of their level of expression in this language. LLMs can be a great equalizer in written documents and allow users to focus on ideas rather than grammar. Some of the tested services specialize in this approach. Both Grammarly and DeepL Pro are excellent at recommending improving grammar, word choice, and phrasing while accounting for contexts. Although at this time, they are less consistent, General models can also be prompted to \prompt{act as a copy editor} and highlight grammatical errors with suggestions for replacement. Some users found that wholesale re-writing of the text by an LLM can change the meaning conveyed. It is thus not advisable to do this, and it is paramount that the writer carefully reads the final text to ensure the ideas are true to intended. Translation services such as DeepL (and Google Translate) can also be helpful for researchers who are more comfortable expressing ideas in their first language. Overall, in this role, LLMs can increase accessibility and ensure that scientific contributions are judged based on merit rather than linguistic proficiency.

Finally, LLMs do not complain when we outsource tedious or repetitive tasks. Such tasks may include formatting citations from one LaTeX template to another, re-formatting a table from HTML to LaTeX or text from Markdown to LateX (e.g., Appendix \ref{sec:survey_questions}), and vice-versa. Reformatting text to fit specific guidelines and to conform to a different template are also trivial tasks that are executed with minimal errors. When significant changes in the text occur (such as decreasing the length of the text to fit a shorter page limit), it is critical to review the final text to ensure the ideas conveyed are the same.

Acknowledging the potential drawback of relying heavily on LLMs for written materials is important. LLMs may inadvertently introduce a standardized or mechanical tone into our writing. Over-reliance on LLMs can diminish your unique writing style and potentially reduce creativity and personalization of communications \citep{doshi2024generativeartificialintelligenceenhances, Chen2023}. Some studies have found that using LLMs reduces overall lexical and content diversity \citep{padmakumar2024does}. Maintaining a balance between utilizing LLMs and preserving your writing flair is crucial. Experiments have shown that humans rely too heavily and gladly on outsourcing decision-making to machines \citep[e.g., Weapons of Math Destruction,][]{ONeil2016}. One approach is to sketch your ideas first and work with the LLM to develop them, rather than outsourcing creativity like "Write me an exoplanet proposal for JWST." It is also important to keep in mind that LLMs carry over the biases of their training datasets. For a further discussion of the ethical issues around LLMs, see Section \ref{sec:ethics}.

\subsection{Reading and Summarizing Papers}\label{subsec:summarizing}

Starting a new project and diving into the literature can be daunting. Traditionally, scientists employ specific patterns of skimming research papers. These approaches vary from person to person due to individual preferences. Different approaches are also employed based on the goal of the search: e.g., looking for a specific technique, a result, a particular value, or a review of a topic.

\begin{figure*}
    \centering
    \includegraphics[width=0.8\textwidth, clip, trim={0.5cm 2cm 0.5cm 2cm}]{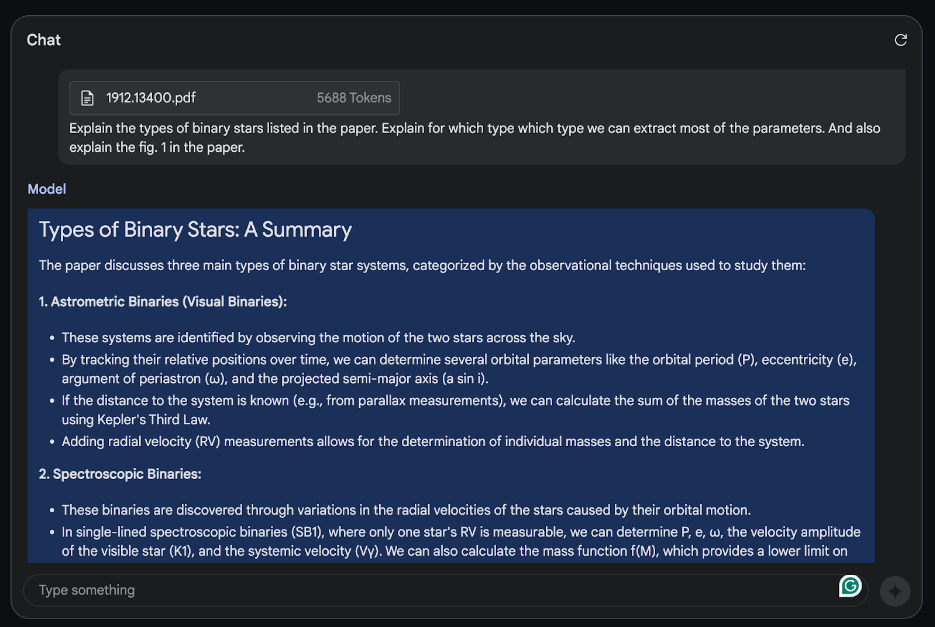}
    \includegraphics[width=0.8\textwidth, clip, trim={0.5cm 3cm 0.5cm 3cm}]{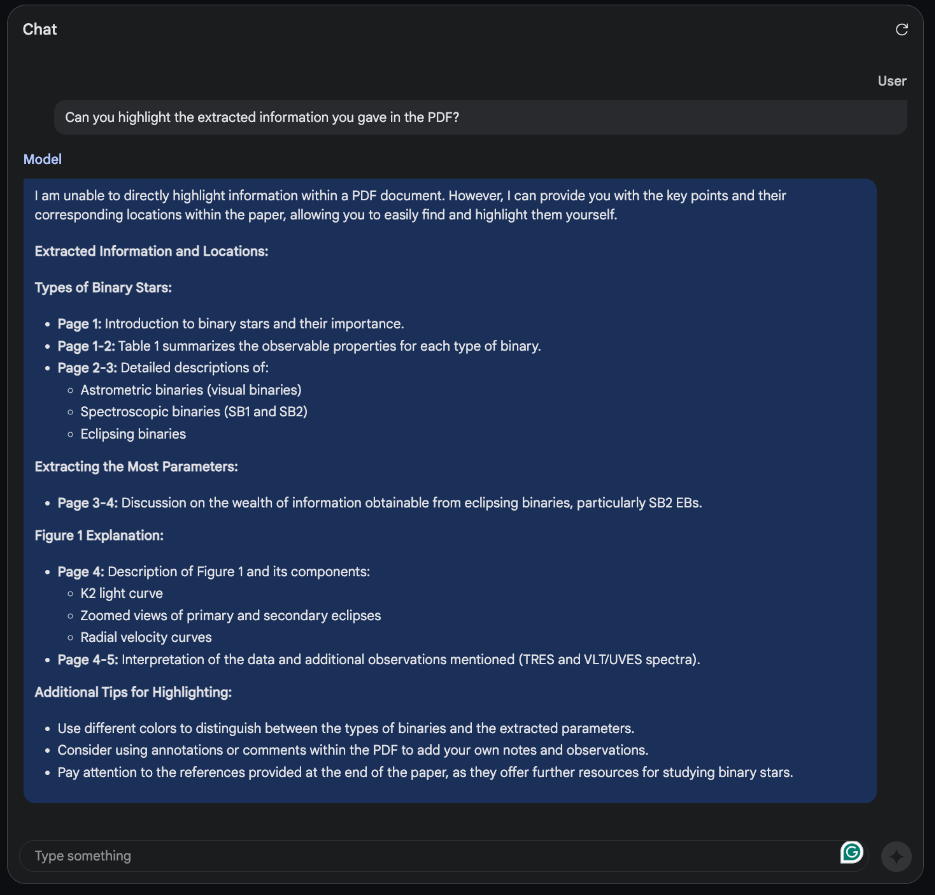}
    \caption{Summary of \citet{southworth2020binarystarscheatsheet} by ChatGPT-4. The chatbot identified the main properties of the stellar binaries discussed in the paper (top) and in-depth content analysis (bottom). It lists sections of the paper focused on different concepts and references page numbers, tables, and figures.}
    \label{fig:sum1}
\end{figure*}

Acting as information retrieval systems, LLMs have the technical capacity to summarize texts and connect scientific papers. This feature can transform how researchers navigate the ever-expanding sea of academic articles and resources \citep[e.g.,][]{iyer2024pathfindersemanticframeworkliterature}.  LLMs can offer a solution by rapidly contextualizing papers and summarising their main ideas concerning a specific question (for instance, \href{talk2arxiv.org}{talk2arxiv.org}). They can extract and list the relevant information and even potentially compare results between publications. In one of the tests for this study, we used the preview model of Gemini 1.5 Pro with a 1M tokens context window in Google AI Studio to examine a PDF file of a paper. Gemini (and other LLMs) allows the functionality to upload PDFs, cutting down the workload of manually putting the paper text. We uploaded an 8-page review article with tables and figures. The Gemini model could extract the information accurately and explain the figures in the paper, albeit not very technically. The chatbot also identified which sections of the paper focused on different concepts by referencing page numbers, tables, and figures (see Fig.\,\ref{fig:sum1}).  

\begin{figure*}
    \centering
    \includegraphics[width=0.8\textwidth]{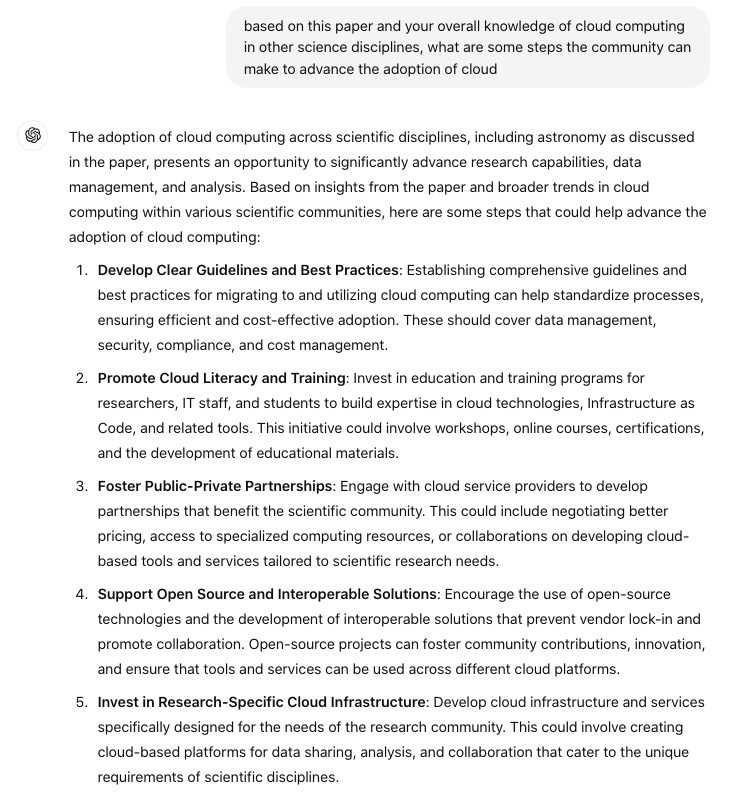}
    \caption{ChatGPT-4 on summarizing \citet{smith}, a US decadal white paper on cloud technologies in science. The chatbot could draw on general knowledge about cloud computing from other scientific disciplines to recommend steps to advance cloud adoption. It also identified challenges not mentioned in the paper (Fig.\,\ref{fig:dec4}).}
    \label{fig:dec3}
\end{figure*}

\begin{figure*}
    \centering
    \includegraphics[width=0.8\textwidth]{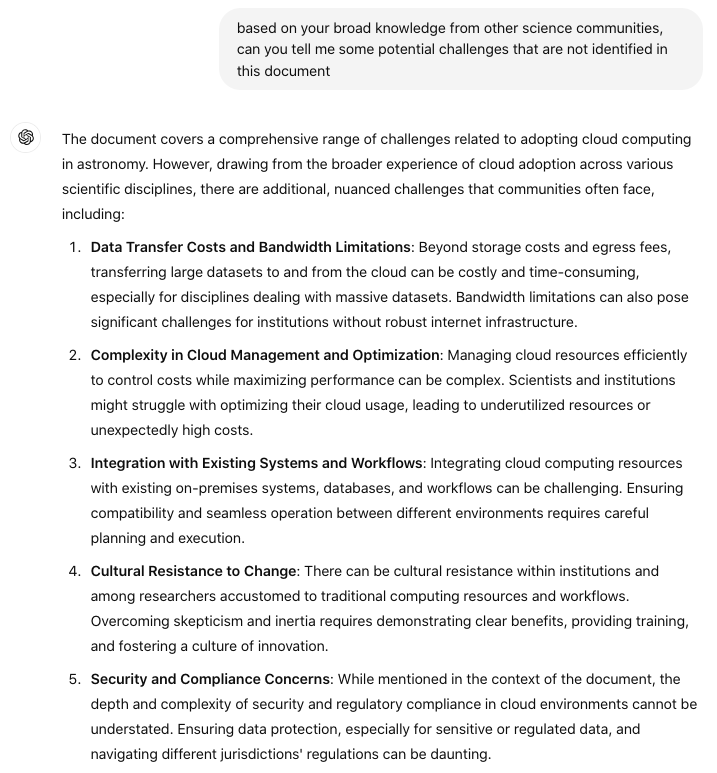}
    \caption{ChatGPT-4 on summarizing \citet{smith}, a US decadal white paper on cloud technologies in science. The chatbot could draw on general knowledge about cloud computing from other scientific disciplines to recommend steps to advance cloud adoption (Fig.\,\ref{fig:dec3}). It also identified challenges not mentioned in the paper.}
    \label{fig:dec4}
\end{figure*}

LLMs also have the potential to help break through disciplinary silos, facilitating the discovery of relevant research across diverse academic disciplines or between academic and industry work. One example of the latter is shown in Figures \ref{fig:dec3} and \ref{fig:dec4}, where we interrogate a US decadal white paper by \citet{smith}. This test was done with ChatGPT-4, where PDFs can be uploaded and parsed. The chatbot could correctly summarize the paper and provide a list of major challenges listed by the authors (response not shown here). But it was also able to draw on general knowledge about cloud computing in other scientific disciplines and make a list of recommendations for advancing cloud adoption (Fig.\,\ref{fig:dec3}) and identify valid challenges not mentioned in the paper (Fig.\,\ref{fig:dec4}). This approach can be useful to authors of papers who can identify gaps in analysis or draw on expertise beyond that of a team. A more advanced take on this approach is to give the LLM a certain persona and ask for feedback based on that persona.

There were occasional glitches, both scientific and technical. Regarding science, LLMs occasionally answered questions incorrectly, especially about niche astronomy topics and physics concepts. As mentioned later, LLMs do not \textit{know} anything, and hallucinations are possible even for the advanced models. Hence, we repeat that it is paramount to check the details of the responses. A technical issue is that the uploaded PDF documents were not always processed in full due to limitations in the context window or issues with document parsing. In particular, they often failed to process tables. The handling of figures depends on the platform and the graphics format (vector vs. raster). Still, the LLMs can usually answer questions regarding the figures based on the context of the caption and text alone. A more general concern is that LLMs may lack a consistent ability to grasp the complexities, uncertainties, and subtle nuances that a \textit{human scientist} would recognize when reading literature. Dependence solely on LLMs for crafting scientific summaries risks producing overly simplistic narratives, neglecting essential arguments and values, and potentially misinterpreting research findings.

While none of the platforms we reviewed are tuned to the specific domain of astronomy and astrophysics, such systems are in development. A model fine-tuned to astronomy, \texttt{Astro-LLAMA}, was published in 2023 \citep{astrollama}, and a prototype of a chatbot platform, \texttt{pathfinder} \citep{pathfinder}, was published during the preparation of this manuscript. Such LLMs can answer much more specific questions with greater accuracy and potentially make connections between new and archival publications while adequately citing the sources. Furthermore, they can lead to an exchange of methods and ideas between fields, which is currently limited due to technical language barriers and jargon, a phenomenon termed "undiscovered public knowledge" by \citet{Swanson1986}. Currently, both \texttt{Astro-LLAMA} and \citep{pathfinder} are only trained on abstract and thus lack knowledge of the more detailed analysis and scientific results, which are only contained in the full-text publications.
However, using the full corpus of astronomical literature may raise legal and ethical concerns (see Sect.\,\ref{sec:ethics}). Software libraries such as \texttt{langchain} enable individual researchers to customize models for use on a limited corpus. Exploring such use cases is beyond the scope of this current paper, but drop-in tools such as Incarmind\footnote{Incarmind: \url{https://github.com/junruxiong/IncarnaMind}} already exist, and we expect such use to become more commonplace soon.

\subsection{Writing Scientific Code}\label{subsec:code}

\begin{figure*}
    \centering
    \includegraphics[width=0.7\textwidth]{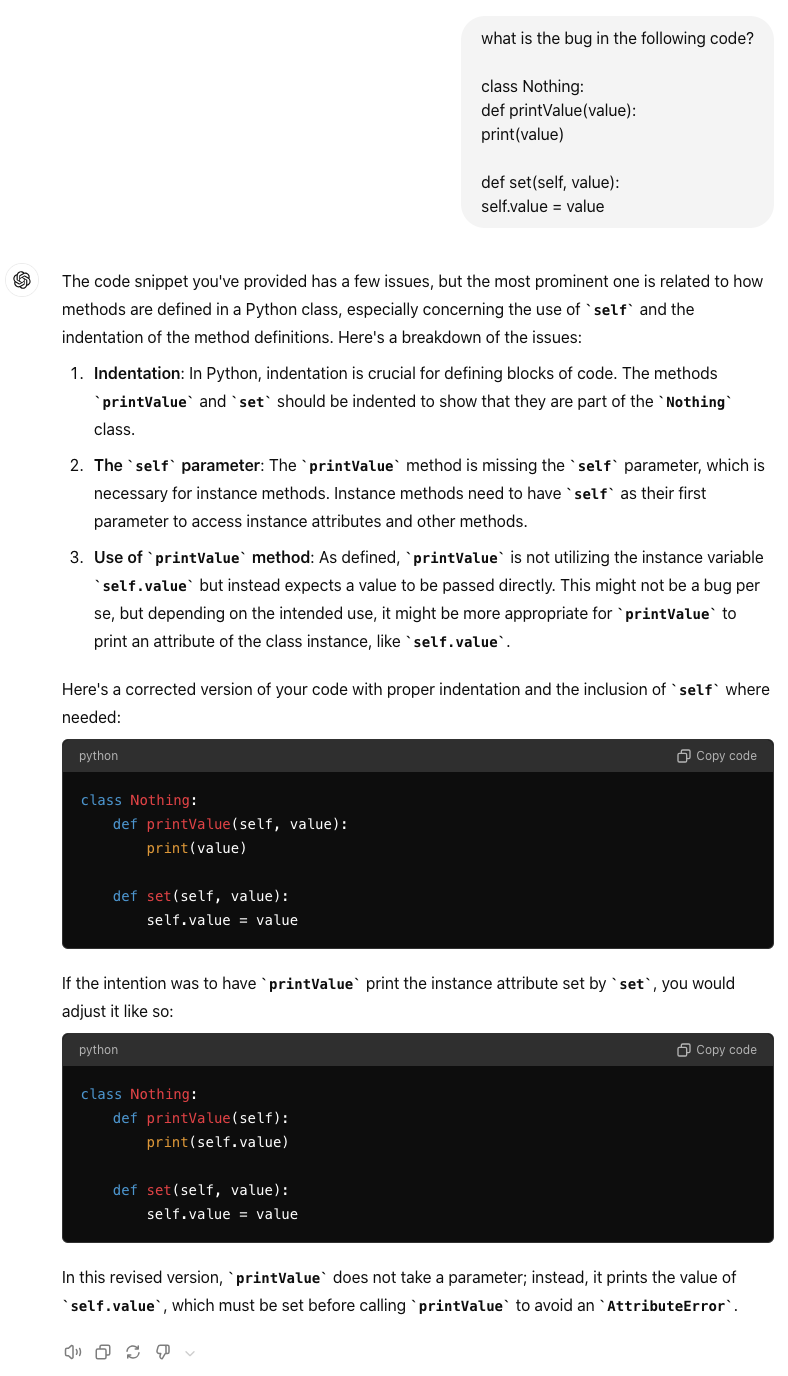}
    \caption{ChatGPT-4 assisting debugging process. This example shows how LLMs can help understand and debug a piece of code. Here, ChatGPT-4 provides corrections and associated explanations.}
    \label{fig:bug}
\end{figure*}

\begin{figure*}
    \centering
    \includegraphics[width=0.8\textwidth]{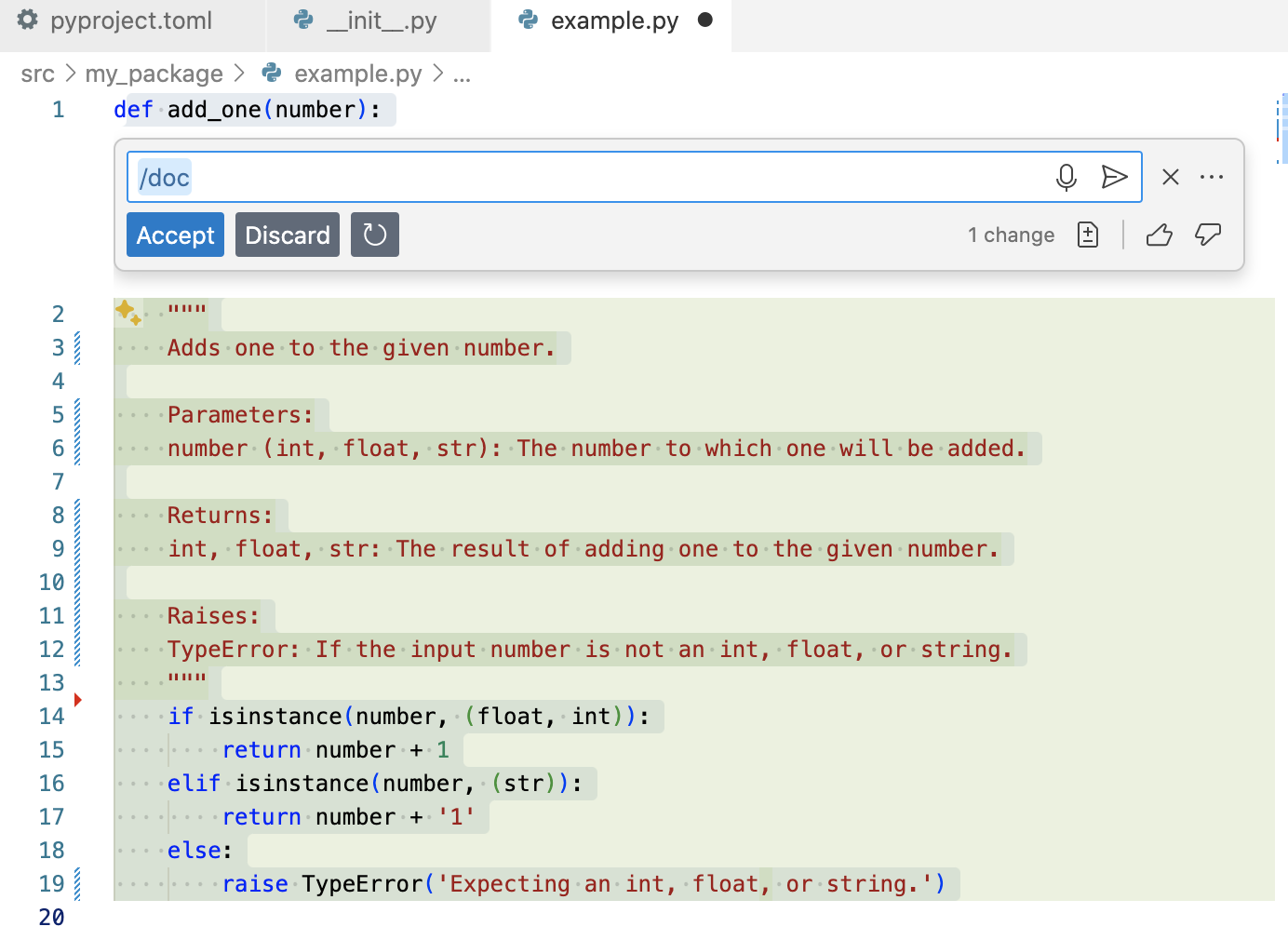}
    \caption{GitHub Copilot could generate accurate documentation with variable typing for an example of a Python function. LLMs can unload the burden of tedious tasks, here documentation, leading to a higher quality standard for source codes.}
    \label{fig:docs}
\end{figure*}

One of the most apparent and exploited use cases for LLMs is their ability to generate and analyze computer code. Programming is a critical skill for research astronomers, as shown by \citet{Momcheva2015}, yet many researchers do not receive training in proper software development practices. Even for those who received training, the wide-ranging tasks involved in scientific programming are constantly changing and frequently very niche. Although we do not think about it this way, coding is a creative task and thus very well-suited to LLMs. Both general-purpose LLMs (Figure \ref{fig:bug}) and specialized ones (such as GitHub Copilot, Figure \ref{fig:docs}) have proven quite adept at programming and have thus become invaluable tools to assist in the workflow of a scientific programmer.\footnote{Scientists also use LaTeX code to write papers; Overleaf provides LLM tools to debug their documents} The tasks they can help with include a range of actions, from explaining code, debugging, or offering simple snippets inside an IDE (Integrated Development Environment) to creating documentation and tests to using chat-based LLMs to create programmatic implementations easily. All of these can drastically improve the speed at which one creates fast, high-quality, scientific code.

In our internal survey, coding assistance is the most common use case for LLMs, with 92\% of respondents. What users find most useful about coding with LLMs is that it helps them complete tasks faster (83\% of responses) and write more code in less time (63\%). They can also identify and fix bugs easier (58\%), improve their coding knowledge (54\%), and learn new coding concepts (54\%). Participants found that LLMs rarely suggest very complex code: 71\% rarely or never received advanced suggestions. Overall, the code contributions are relatively minor, with 42\% of the respondents saying that LLMs have contributed 1-20\% of their code and only 8\% saying that LLMs have contributed more than 80\% to their code. Users are more satisfied with the accuracy of the coding prompts than they were with their writing prompts: 72\% of respondents found that more than 60\% of their coding prompts were successful. A handful of respondents have tried translating code from one programming language (Julia, IDL, Fortran) to another (Python) and have found varying degrees of accuracy and satisfaction.

Arguably, one of the best tools for auto-completion-assisted coding is Github Copilot, as its context window is large enough to include multiple files at once. This feature is helpful with projects or packages where the quality of the suggestions improves dramatically when the LLM is aware of all the code in the project. In addition, GitHub Copilot includes a chat window in some IDEs, such as Visual Studio Code (VS Code, Figure \ref{fig:docs}), making it even more convenient. Codeium represents a free alternative to GitHub Copilot that works similarly but provides a smaller context window and no chat capability (in the free individual version). Both companies support extensions for many editors and platforms.

Chatbot platforms, such as ChatGPT3.5/4, Gemini, and Claude, have the advantage that even less experienced programmers can directly formulate their request in plain language ( English or any other language), and the LLM will provide a snippet of code in the requested language. The code can be copy-pasted from and to any terminal or editor. Team participants who tested this functionality frequently queried solutions to syntax errors but found the provided code helpful nonetheless. LLMs can also assist programmers by doing arduous or repetitive tasks (e.g., regex string, parsing, code translations from one language to another, etc.) for them, which are often hard to understand for humans, but the LLMs excel in.

To assess the behavior of chatbots on more complex programming tasks, we tested ChatGPT-3.5 and ChatGPT-4 on a more advanced coding example: "\prompt{Please create a simple beta-Variational Autoencoder model implemented using PyTorch. The model inputs are a set of images and their sky coordinates. The output should be a bijective mapping by coordinates w.r.t the distribution of images in the latent space.}" In addition to being complex (several interconnected functions are required), this is not a very common piece of code. The outputs from the LLMs differed: ChatGPT-4 described the model construction in detail, including data preprocessing, network architecture, loss function, and training loop, while ChatGPT-3.5 provided more detailed code and less conceptual explanation. While neither chatbot provided a perfect working example, both gave the user an adequate implementation of the base model as a starting point. 

As with other use cases, users can receive significantly different responses depending on the wording of the prompt (also see Sect.\,\ref{sec:prompting}). Some participants found it useful to include the user expertise level in the prompt - for example, requesting code for a beginner programmer frequently resulted in a response with more code comments and a more detailed explanation of the code.  In a more complex code problem, it was useful to start with a simple question and then build upon it with additional questions or steps. Regardless of writing a good prompt (see Sect.\,\ref{sec:prompting}), the accuracy of a generated code depends on the actual task and the libraries required to accomplish it - for rare and particular tasks requiring singular niche libraries there are often not enough training materials to provide reliable responses (e.g., \prompt{How can I use Pyphot to get the photometry of a few spectra through the Gaia passbands?}). The code presented in these chats is often flawed to a certain degree, with syntax/API errors being quite common. Users can correct the error(s) themselves, depending on their expertise level, or provide repeated input to incite the LLM to fix its errors (e.g., copy-paste the error and ask the LLM to troubleshoot). Still, after too many iterations, the context window can saturate (especially using free tools), and the LLM may trap itself in a repeating cycle. 

Overall, LLMs can only (at this time) create straightforward scripts directly and without error but often provide a perfect starting point for more complex tasks for any programmer to improve upon. They can give ideas (such as suggesting an approach or even specific packages that could be helpful) that might be outside the programmer's direct scope/knowledge and thus lead to either a learning effect and/or improved code.

It is essential to mention that the free versions of these tools, like Codeium, ChatGPT3.5, and others, are generally sufficient to assist with relatively simple scientific coding. Paid services (e.g., GitHub Copilot) provide more specific answers as they handle a more extensive project context and can understand more complex tasks. In the case of the chat LLMs, the features vary with the versions, like web-based searches, thus more up-to-date information, which is appreciated to avoid outdated and possibly deprecated code snippets.

Finally, regardless of which model, it is imperative to be knowledgeable enough to understand if the LLM is on the wrong path or producing direct errors, referencing outdated/deprecated packages, having logical flaws, and showing convoluted solutions. Lack of understanding is risky and incredibly impactful if inexperienced programmers are using the output of LLMs to improve their coding.

\subsection{LLMs as Data Analysis Assistants}\label{subsec:data_analysis}

Beyond software development, the proliferation of AI tools on platforms like Deepnote, Jupyter, and VScode is revolutionizing how scientists approach data analysis as a whole. AI tools with large context windows can access the entirety of a project, gaining a comprehensive understanding of the code, data warehouses, and metadata. This extensive context awareness enables them to provide precise and customized assistance tailored to the specific work undertaken.

The benefits of using AI tools for data analysis are multifaceted. First, they significantly reduce the time required for data exploration and analysis. For example, they allow the scientist to focus on the question in natural language, such as "\prompt{Filter the data only to include entries with proper motions and parallaxes}" and let the AI create the code, rather than have to parse through the (frequently complex) database schema documentation. Second, by automating routine tasks and providing real-time guidance, AI tools could enable scientists to focus on more complex and creative aspects of their work. 

\begin{figure*}
    \centering
    \includegraphics[width=0.9\textwidth]{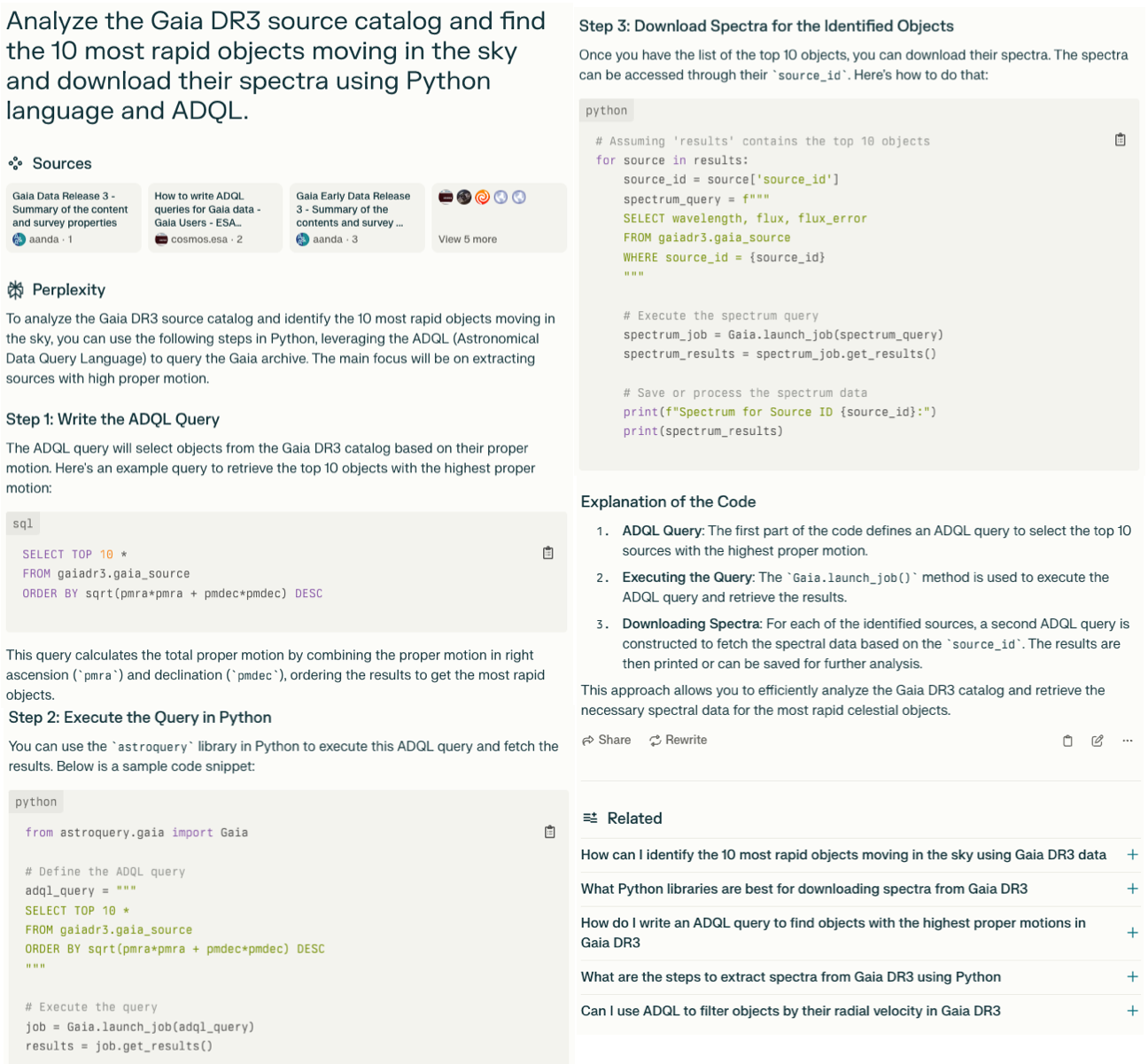}
    \caption{Perplexity on querying the Gaia Archive to find the fastest moving objects. This example illustrates the understanding of the question and the production of a usable output.}
    \label{fig:gaia}
\end{figure*}

Crucially, these AI tools eliminate the need to start from scratch. Scientists can prompt the tool with a specific task or query, such as \textit{"Analyze the Gaia DR3 source catalog and find the 10 most rapid objects moving in the sky and download their spectra using Python language and ADQL."} The tool then leverages its understanding of the project to define the relevant query for online services, streamlining the data analysis process. Figure \ref{fig:gaia} shows Perplexity's response to this query, which details the reasoning and explains the proposed Python code. It also provides the relevant part for downloading the spectra, which the popular Python libraries do not cover. Adding radial velocities to the movement calculations is subtle because it would require some distance estimates. Perplexity provides this limitation if prompted: \prompt{Without specific distance information, we can't directly convert radial velocities to mas/yr}. However, as we mentioned for other tasks, the answer varies between LLMs, and it may vary with time for a single LLM. Regardless, they provide a robust starting point for the scientist to refine.

AI tools can lower the barrier to analyzing complex datasets and potentially fill the gaps in programming knowledge. However, AI tools are not a replacement for human expertise. Instead, they serve as powerful assistants that augment and enhance the capabilities of scientists. Most code outputs or data analysis results require adjustments, enhancements, or refinements. Deepnote.ai published a 12\% acceptance rate \footnote{Impact of Deepnote Copilot \href{https://deepnote.com/blog/deepnote-ai-so-whats-the-impact-part-1}{blog post at deepnote.com}
} of AI-generated outputs on their platform. A small change in the AI prompt could often lead to different outcomes (see \ref{sec:prompting}).

\subsection{Image Generation for Outreach}\label{subsec:outreach}

When communicating astrophysical research to the general public, it is often important to supplement scientific material with visually impressive, easy-to-understand images and videos to capture the audience’s attention and clarify key points. Major differences emerge when comparing plots for a scientific audience to outreach images used in popular presentations. First, “mathematical-looking” plots, labels, and similar elements should be reduced to a minimum. This avoids confusing non-experts with unfamiliar concepts like logarithmic scaling and also helps retain attention, as many people are "turned off" by overly technical or mathematical details. Second, it is sometimes necessary to sacrifice accuracy for simplicity. In line with the first point, focusing on a single, basic idea is often more effective than presenting nuances meant for a more specialized audience. Third, it can be beneficial to prioritize visual coherence or "beauty" over precision. The goal is to capture attention, and visually striking schematics or animations will aid in doing so. As such, a “good outreach image” fulfills the following criteria: (1) Easy to parse and understand the key point, even without scientific background; (2) Visually impressive and coherent; (3) Scientifically sound, even if minor or technical details can be omitted.

However, most astronomers have little to no experience in areas such as computer animation or digital art. They are rarely tasked with making the images used with, for example, press releases. Regardless, these images and animations can be helpful tools when communicating scientific findings, particularly for a lay audience, for the above reasons. 

Generative AI, and more specifically, image generation models (such as DALL-E), can assist in scenarios with limited access to artists or animators. Even in these scenarios, it is important to avoid using images and videos at the risk of communicating incorrect science. Thus, it is imperative to assess the generated images according to the criteria outlined above.

In this study, we attempted to generate a handful of outreach images for some astrophysical concepts using BingAI's {\it Creator} and ChatGPT, both of which use DALL-E as the underlying image generator. The main difference in these models is their specific conversion or translation of the user prompts to DALL-E prompts fed to the image generator, which occurs “under the hood.” For each experiment, we iterated several times to optimize the outcome, updating the user prompt to suit our needs. For brevity, we only present one such case here, generated using ChatGPT to create a visually appealing picture of a stellar cluster with tidal tails. This experience poses an interesting example, as there are few images of this astronomical concept aimed at a lay audience. Further, it combines multiple astrophysical aspects, thus creating a challenging prompt for the LLM. Table \ref{tab:outreach} illustrates the generation process through multiple iterations.

The LLM struggles to encode the astronomical jargon into a scientifically sound image. Inherently, the challenge here is that "Cluster" and "tidal tails" have specific meanings in an astrophysical context but also connect with more popular contexts outside astronomy. It is perhaps not particularly surprising that without reference image prompting, the generated images consistently show "clusters" or blobs of stars within a larger structure. Further, the use of "tides" likely evoked a relation to oceanic tides, and thus wave-like structures were generated. An attempt to first get the "cluster" portion right, while leaving out the "tides", resulted, again, in blobs of stars, now without the wave-like structures. After showing an image of a globular cluster with tidal tails to the LLM, it managed to move away from repeating the same structure but produced something more akin to a spiral galaxy. Asking it to create something more aligned with the reference image again prompted the blob structure.

This example demonstrates that context is essential when parsing prompts for image generation in LLMs. One must word prompts carefully and precisely, ideally omitting jargon and especially polysemes, i.e., expressions that have more widespread sense. Reference images can be helpful but do not guarantee a desirable outcome. Great care has to be taken when generating images for outreach with tools like ChatGPT to avoid confusing a lay audience with potentially erroneous images.

At this present stage, our experience shows that AI tools could help with simple illustrations but not with complex concepts. For the next few years, artists or animators will continue to be the path of choice for creating effective illustrations.

\section{Discussion}\label{sec:discussion}

Following the presentation of our findings, this section summarizes the main benefits and limitations, discusses legal and ethical concerns, and discusses general tips for making the most of using LLMs for research.

\subsection{Benefits of LLMs in Research}

LLMs have many and varied applications to tasks carried out by researchers, as demonstrated in the use cases discussed in Section \ref{sec:use-cases}. 

LLMs can assist in communication tasks: drafting emails, letters, papers, and proposals. They can also help in editing and proofreading academic manuscripts, formatting citations and bibliographies, and generating content for teaching. They can summarize individual or multiple documents, identify gaps in analysis, and create summaries of existing research and literature. They can also identify trends in the field, making them especially useful for literature review. LLMs can generate and analyze code, assist in debugging and refactoring code, and help in code documentation and explanation. LLMs can process and analyze large datasets, extracting key insights and information and aiding decision-making processes.

Integrating Large Language Models in research workflows offers numerous advantages, from enhancing communication and facilitating literature reviews to streamlining data analysis and supporting software development. LLMs have the capacity to significantly reduce the time required for data exploration and analysis. For example, they allow the scientists to focus on the question in natural language rather than parse through (frequently complex) sources of information. By automating routine tasks and providing real-time guidance, AI tools could enable scientists to focus on more complex and creative aspects of their work. As researchers increasingly adopt these tools, the potential for significant productivity gains becomes evident. In our survey, 83\% of respondents think that LLMs will become an integral part of science, which likely translates the versatility of AI to be applied across various tasks in academia.  

Research is inherently a process of complex knowledge-based tasks, which rely on a pool of experts with significantly varying abilities and knowledge. People who would likely get the biggest boost out of this technology are those with the lowest initial ability in a given task, especially students. Still, even those with the highest level of skill are likely to benefit in terms of productivity and efficiency \citep[pages 156-157, ][]{Mollick2024}. At the same time, subject expertise will become not less but more important. Expertise requires knowledge of facts, extensive practice, critical thinking, problem-solving, and the ability to think through problems and troubleshoot. Being the responsible "human is the loop" requires us to be engaged in this collaboration with LLMs but also allows us to learn from them; it fosters a sense of responsibility and accountability \citep[page 54, ][]{Mollick2024}. Economists predict that AI is unlikely in the next several years to cause dramatic changes in most jobs  \citep{Ilzetzki2023}, and we believe this to be true about the work of researchers in astronomy as well. As demonstrated by our survey (Sect.~\ref{sec:survey_questions}), most participants reported relatively minor contributions to their code. Instead, we will likely experience many small workflow changes that will improve efficiency, and we will delegate some tedious and/or repetitive tasks to LLMs and use them as tools to enhance our creativity.

\subsection{Limitations and Responsible Use}\label{sec:limitations}

While LLMs offer remarkable potential, we must understand their limitations and exercise caution in their application. Our survey raised concerns about these limitations, with 25\% of respondents expecting that LLMs will have a negative influence on science overall.

Hallucinations are one of the most frequently encountered drawbacks of LLMs \citep{fecher2023, Birhane2023}. Hallucinations are instances where a language model produces false or misleading information that seems plausible but is not grounded in reality. This is an issue we caution against in each use case above. The most commonly known examples are of LLMs making up citations in documents or research papers. Newer and larger LLMs hallucinate significantly less than older versions. In practice, the reduction of hallucination is primarily produced by mitigation mechanisms (e.g., Retrieval-Augmented Generation, RAG, post-processing techniques, user feedback), not by changing the fundamental algorithms. For example, \citet{hallucinations} found that while ChatGPT-3.5 made up fully 98\% of citations, ChatGPT-4 only made up 20\% of them. Models connected to the internet are also more likely to link to real resources on the internet but are not entirely devoid of this issue.

It is unclear if hallucinations are a solvable problem. LLMs are ultimately elaborate auto-complete engines based on statistical patterns with no concept of facts, truth, originality, or meaning. They do not actually know anything: if you ask an LLM for a citation, a quote, or a number, it will generate the answer based on the statistical connections in the model rather than retrieve it from memory. For this reason, use cases where we ask LLMs for specific numbers and figures are inappropriate use of this technology and not playing to its strengths - we are trying to expose rather than explore the technology \citep{Mollick2024}.  

But beyond the obvious errors, such as numbers and citations, there is also a vast space of more subtle inaccuracies, which we need to be cautious about. For instance, using LLMs to edit or paraphrase text can change the ideas and meaning of a text; in summarizing papers, LLMs can over-simplify or misinterpret research findings; in code, LLMs can change the flow of an algorithm or the indentation of code. These are all due to the LLMs' fundamental lack of contextual understanding and reasoning capabilities \citep{Birhane2023}. While these models can identify patterns and generate plausible text, they may struggle to grasp the nuances and complexities of scientific concepts. Astronomers should view LLMs as tools to augment their research efforts rather than replacements for human expertise and critical thinking; they must critically assess the outputs of LLMs and verify their accuracy against established literature and empirical evidence.

We purposefully do not explore in detail the use cases of peer review and assessment of papers, proposals, job applications, etc. Even though these are some of the most time-intensive tasks and obvious candidates for LLM automation \citep[e.g.][]{jerabkova2024scientifictextanalysisrobots}, we believe these are tasks that fundamentally should not be outsourced to machines. Using LLMs in tasks that relate to the allocation of resources and assessment of performance raises several ethical and legal issues (see Section \ref{sec:ethics}), and it also endangers our trust in the fairness of the process \citep{Birhane2023}. While we can hold people responsible for misinterpreting a proposal or an article, it is unclear who should be held responsible if the machine makes an error. 

Likely, the most fundamental concern that underlies the debate of LLMs in research writing is whose ideas are those. We have so far existed in a world where (in academic writing) written materials capture our own ideas (and those of our collaborators) and reflect our thoughts. And in this world, using LLMs for writing "feels" different than, for example, coding. While we consider software as a means to an end (data reduction, analysis, plots), the interpretation used to come from the human scientist. Writing with an LLM may feel intellectually dishonest because it is unclear that the ideas in the final text are those of the authors. Research has shown that the process of writing helps us think \citep[e.g., ][ and references therein]{writingandthinking}, and it is not clear to what extent this holds for the LLM-enhanced writer.

\subsection{Ethical and Legal Concerns}\label{sec:ethics}

Many ethical and legal concerns have been raised about using LLMs, both broadly within society and within astronomy specifically. In this section, we summarize the main concerns. We must recall that we are not experts in legal propriety and ethical research.

\subsubsection{General Concerns}

One major issue raised against tech companies is using copyrighted materials to train LLMs in the spirit of "better ask for forgiveness than for permission." LLMs are unlikely to reproduce text wholesale since they do not save the text itself but only the statistical weights across many texts. Only in cases where the text is very common \citep[e.g., the Gettysburg address, Alice in Wonderland, ][]{speakmemory} do LLMs auto-complete sentences correctly. Copyright laws in Europe and the USA are still unclear on whether this is an infringement, while in contrast, the Japanese law explicitly includes provisions accommodating AI training \citep{japaneselawcopyright}. While not outright illegal, the practice is considered unethical by many. Many scientists may believe that using the product makes them complicit in copyright infringement and choose to avoid using these services. 

A specific example of this issue in the context of astronomy is the advent of astro-specific LLMs such as {\texttt{pathfinder}} \citep{pathfinder} and \texttt{Astro-LLAMA} \citep{astrollama}. While these models are currently trained only on the abstracts of papers, \citet{pathfinder} mention the possibility of training on the full text of papers, and \citet{astrollama} explicitly say that they are in the process of ingesting the full \LaTeX versions of papers. But both \texttt{arXiv} publications and papers linked to from the Astronomical Data System (ADS) have licenses, which in many cases prohibit reuse and reproduction without the explicit permission of the authors. It is impractical to obtain permission from every author, making models less powerful to only use manuscripts with Creative Commons licenses. However, ignoring the licenses on astronomy papers is just as illegal and unethical as ignoring the licenses on New York Times articles. 

It is important also to consider the ownership of intellectual property of materials given to and produced by LLMs for individual users. The data we provide to an LLM, such as the text of a proposal or a paper, may not be immediately incorporated into the model's training corpus but could eventually become part of it. Although it is highly unlikely that the ideas from one user will be presented verbatim to another, they are not discarded. This may be especially critical for users with sensitive or proprietary information, such as writing code for hardware under export controls or the International Traffic in Arms Regulations (ITAR). In some cases, platforms offer paid access with a promise that personal information will not be used in training, but ultimately, users have very little control over their information. It is also unclear if text generated by LLMs is protected by copyright and who owns the copyright. In some cases (e.g., OpenAI), the service agreement grants the copyright to the user, but the issue is still open from a legal standpoint. 

Another ethical concern about LLMs is that AI companies hire low-paid English-speaking workers in developing countries to do the reinforcement learning through human feedback, a process that trains models to avoid unethical or dangerous responses, subjecting these employees to unethical and dangerous responses. Many researchers may not be aware of this part of the training process and may object to their use based on refusing to be a part of an exploitative labor system (however, this system is not unique to LLMs, and similar systems underlie the content moderation of many social networks). 

The source of materials for training LLMs -- largely the internet -- is also not inherently unbiased. Many studies have shown that there are gender differences in both participation and representation on the internet \citep[e.g., Wikipedia, StackOverflow, GitHub;][]{Noei2021, May2019, Stephens2013, wikipedia, github}. Additional cultural biases are introduced (or amplified) by the fact that data are dominated, if not limited to, English-language documents. An exploration of image generation by Stable Diffusion by \citet{Nicoletti2023} revealed strong gender and race stereotypes, depicting higher-paying jobs as predominantly male and white. The biases were stronger than in real life: while 39\% of doctors in the US are women, only 7\% of generated images of doctors were female.  The emerging field of "machine psychology" is only just starting to understand the behavior and reasoning abilities of LLMs. Still, early work has shown that they exhibit at least some of the same cognitive biases as humans \citep{llmbias}. Models can bring their learned biases to recommendation letters, job applications, performance reviews, etc., in the same way as humans.

A final general concern is the environmental impact of LLMs and image generation models both during their training and subsequent use. For example, training a single large-scale model can emit as much CO$_2$ as several cars over their entire lifespans \citep[e.g., ][]{nlpcar}. The frequent updates and fine-tuning of models, as well as the ongoing use of these models, also contribute to their environmental impact due to the continuous demand for powerful servers and cooling systems in data centers \citep[e.g., ][]{ai_env}. Image generation models are especially power-intensive - one generated image can use as much power as fully charging a smartphone. In contrast, one text generation is equivalent to 0.0016\% of a smartphone charge \citep{Piper2023}. As these technologies become more prevalent, we need to account for their contribution to the CO$_2$ footprint of research activities. 

\subsubsection{Research-specific Concerns}\label{sec:research-specific-concerns}

Within the broader scientific community, LLMs have triggered spirited debates with opinions ranging from an open embrace to a complete rejection of AI tools. 
\citet{Dwivedi2023} collected opinions within the first couple of months after the release of ChatGPT and demonstrated the range of discourse, from effectively equating LLMs with plagiarism and calling for a ban (Contribution 29 in the cited paper) to frank discussions of the ethical quandaries of what research means (Contribution 27) to the open acknowledgment that certain contributions have been primarily written by ChatGPT itself (Contribution 18). 

One major issue in this discussion is whether using LLMs in research is injurious to the scientific process. Philosophers of science are only now starting to consider the implications of LLMs for science. One approach is to remember that science is fundamentally a human enterprise determined by the historical, social, and cultural context in which it happens, and our questions, methods, and analyses are influenced by our backgrounds \citep{Birhane2023}. LLMs are just tools that have none of these contexts and motivations. Human scientists are fully responsible for the work they produce. But it may be difficult to determine whose ideas are those presented in the final work -- do they belong to the scientist or the training set \citep[][Contribution 25]{Dwivedi2023}? Research has shown that the process of writing helps us think \citep[e.g., ][ and references therein]{writingandthinking}, and it is not clear to what extent this holds when using LLMs to assist human writers. The key point is that the ideas in the written text and materials should reflect the authors' thoughts, which may be challenging when collaborating with LLMs \citep{vishniac2023}. Possibly, this represents the crux of why we think using LLMs for writing "feels" different than for coding. While we consider software a means to an end (data reduction, analysis, plots), the interpretation should come from humans. Writing with an LLM may feel intellectually dishonest because it is unclear that the ideas in the final text are those of the authors. Astronomers are certainly not the only ones facing these new issues. Scientists from all disciplines and philosophers of science are only now starting to struggle with such questions, and there is an ongoing open debate across academia on the ethical use of LLMs. We present an abbreviated slice of this discussion here, and we urge members of the astronomical community to be engaged and informed participants in the debate. 

First and foremost, intellectual ownership represents a major issue for the integrity of the publication process.  Currently, several scientific journals have published policies regarding the use of LLMs in writing research manuscripts. Most relevantly for astronomers, in March of 2023, the editor-in-chief for the AAS journals provided a clear guideline for the use of LLMs in scientific manuscripts submitted to the AAS \citep{vishniac2023}: (1) the authors of the manuscript are solely responsible for its content, and (2) LLMs should be cited as software. The Monthly Notices of the Royal Astronomical Society (MNRAS)\footnote{https://academic.oup.com/mnras/pages/general\_instructions} has issued a similar statement as has Nature\footnote{https://www.nature.com/nature-portfolio/editorial-policies/ai}, additionally requesting that peer reviewers avoid the use of LLMs due to privacy concerns. Other major journals have adopted similar stances, and only rarely are there policies that completely ban LLM use (e.g., ICML\footnote{https://icml.cc/Conferences/2023/llm-policy}). 

Yet, the recommendation across all journals that using LLMs should be acknowledged does not seem to be abided by. As of August 2024, there are only 135 mentions of "ChatGPT" in the full text of articles of the ADS astronomy collection since the start of 2023, while {\texttt astropy} is mentioned 4,957 times. Further, as shown in our survey (see Section \ref{sec:survey_questions}), even though 78\% of participants agree that it is necessary to credit LLMs in papers, only 28\% say they do so every time they use them. A more quantitative analysis has been done by \citet{astarita2024}, who show a statistically significant increase of words commonly used by LLMs in the astronomy literature since 2022 \citep[see also][]{geng2024}. There is a widespread lack of compliance with journal guidelines, and using LLMs likely extends beyond journal articles. This lack of compliance is likely due to what \citet{PorsdamMann2023} put in the context of a credit-blame asymmetry: generative AI diminishes the apparent effort of the authors (i.e., gives them less credit for their work). Still, it increases their responsibility because LLMs are error-prone. Within this context, authors are not incentivized to report using LLMs because doing so will result in their work being judged more harshly. In this regard, LLMs are different than other software tools. As a result, the recommendation to cite LLMs as software is at odds with the psychology around LLM usage. 

Of course, it is also impossible to enforce the reporting recommendations because it is practically impossible to tell if a specific text is fully or in part written by an LLM \citep{detection1}. Prompting strategies and small changes to the text can evade existing detection methods. Worse than that, LLM detectors have high false positive rates for non-native speakers and writers with limited vocabulary \citep{detection2}. Therefore, providing guidelines on the appropriate and inappropriate use of LLMs is especially critical. 

Second, there are real new obstacles that journal publishing is likely facing. Irresponsible authors may submit text with fake citations, which puts the burden on the reviewers or copy editors to rectify the issue later. Fake citations may also be seen as "the canary in the coal mine" - a sign that there may also be factual errors in the text as the authors have used LLMs and not bothered to check the text they produced. Automated citation checking at submission has been implemented by some journals (e.g., JOSS) and can/should be done before review to flag such issues. While there could be automated mechanisms to check citations, generated reviews are challenging to detect, as mentioned above. A more insidious problem is that authors can generate and submit multiple versions of the same paper to the same or different journals to boost their publication records.  This can become a major issue, wasting the time of editors and reviewers while clogging the system with "spam" submissions. However, the ethics standards of most journals already cover those issues. For example, the AAS Journals Professional and Ethical Standards state\footnote{https://journals.aas.org/professional-and-ethical-standards-for-the-aas-journals/} that authors submitting their manuscript certify that it "has not been previously published elsewhere and is not being considered for publication elsewhere in any form." A system that enforces this rule (and creates consequences for its violation) may be able to safeguard journals from such spam publications. The other side of the coin is the reviewing process being hampered by generated reports. Within the context of intellectual property rights discussed above, this is a more clear-cut issue: the authors hold the copyright of their work, and therefore, third parties (i.e., reviewers) should not be allowed to share these texts with LLMs where they could potentially become part of future training sets.

Finally, there are rising concerns about using LLMs to accelerate tedious tasks critical to (astronomy) research. One of the most labor-intensive processes in the astronomical community is telescope proposal review. Therefore, it is unsurprising that there is a lot of interest in the performance of LLMs for this task. Our team was curious to test this performance. For this work, we used ChatGPT-4 to assess five ESO proposals from previous cycles. We provided the scientific rationale of the proposals as well as the review guidelines and the grading scale ranging from 1 (outstanding, ground-breaking) to 5 (unsuitable) to the LLM. Human reviewers assessed these proposals with grades between 1.8 and 3.8, according to the ESO rubric. In its first response, ChatGPT-4 dissected the proposals by answering each of the 8 questions provided in the guidelines\footnote{ESO guidelines for distributed peer review are available at \\ https://www.eso.org/sci/observing/phase1/distributed-peer-review.html}. It correctly identified the relevant points in the proposals to address these questions. However, it ranked all 4 proposals as excellent, with a grade of 1.5. Subsequent prompts to "compare the proposals," to "rank the proposals," and to "use the full grading scale" resulted in the model randomly re-ordering the proposals. The final grades still clustered between 2.0 and 2.75. This small experiment shows that ChatGPT-4 can be persuaded when assigning scores and that it struggles to keep consistency between grades. 

Furthermore, ChatGPT-4 cannot detect weaknesses in the proposals, tending to assign excellent grades in all cases. \citet{jerabkova2024scientifictextanalysisrobots} proceeded to a more in-depth experiment with another set of five proposals and reached similar conclusions. This behavior is also present in humans, although less common with increasing experience. The caveat of both studies remains the limited sample of five proposals. For now, LLMs can help extract and summarize information from the proposal. Still, the downside is that they remove important information for the reviewer, such as clarity in presenting the case and objectives. Finally, LLMs cannot properly assess the potential impact that the proposed observations could have in their fields, which is only possible for a trained astronomer with real knowledge and critical thinking. 

Just because LLMs can do something (such as a proposal or paper review) does not mean we should delegate the task to them. As discussed by \citet{Ntampaka2022}, tasks that relate to the allocation of resources, whether at the PI level \citep[e.g., grants, observing time proposals, ][]{jerabkova2024scientifictextanalysisrobots} or at the community level \citep[e.g., decadal recommendations, ][]{Thronson}, should not be delegated to our robot overlords even when it is extremely tempting to do so. The same is true for decisions that involve hiring and promotions. In these areas, there are many examples where machines internalize and amplify existing biases. Multiple studies have shown that when humans supplement decision-making with algorithms, they are much more likely to stop paying attention and outsource the decision-making to the algorithm \citep[a process called "falling asleep at the wheel," ][]{fallingasleep} and such decisions are (at least for now) perceived as less fair \citep{NEWMAN2020149}. Even if one can build models where metrics adequately address fairness and ethics, dangers still exist. Experiments such as the one in the previous paragraph, those presented by \citet{jerabkova2024scientifictextanalysisrobots} and \citet{Thronson} are interesting, but systems like these should be avoided in practice as they can entrench biases and undermine the fairness and integrity of these decisions. Authors and organizations that pursue the creation of such systems should deeply consider the ethics of such experiments.

\subsubsection{Does the cost of AI generate inequalities in academia?}

As AI tools offer various features that can enhance productivity and efficiency in scientific research, they could be a source of inequity. When only a portion of society has access to a piece of technology, we risk creating a digital divide where the have-nots are disadvantaged. The cost associated with premium versions of these tools can create disparities in access and usage among researchers.

Most platforms currently provide a free tier, which limits usage regarding the number of queries, general access to the latest models (usually more accurate), or even advanced tasks such as web queries and code testing. The free tiers of both Perplexity and ChatGPT do not support advanced data analysis features, for instance. Users on paid tiers can upload files (like PDFs) and utilize tools to create interactive charts and detailed data visualizations, which are essential for in-depth analysis in scientific research. Some platforms provide advanced or full features with education licenses (e.g., GitHub Copilot with a free tier for students and educators); others (e.g., DeepNote) do not provide complex data analysis features on their free or education tiers; and many do not offer education discounts at all. As our survey in Section \ref{sec:survey-results} highlights, more than half of the participants recognize that paid services are superior and express interest in paid access.

It is also important to clarify that free access means that the user is the product: the data from free users are used to test the model and potentially train its next generation. One argument for paid subscriptions is that their service agreements sometimes (but not always) include a clause that user data will not be used as future training material. However, users are unlikely to have any control over the use of their data and cannot guard against future changes in the service agreement (also see Section \ref{sec:ethics}).

Open-source models offer a different approach (e.g., the Llama community\footnote{\url{https://www.llamaindex.ai/}}). Some are available for download, and users can run them locally, frequently even on a laptop, and in some cases with a convenient in-browser interface, even if they are not as feature-full as some commercial services. While these are more elaborate to set up, they allow users to bypass some of the concerns or limitations discussed above. Encouraging the development and use of open-source AI tools could also help mitigate inequalities.

The cost of AI tools in academic research has the potential to create additional inequalities between institutions: those with larger budgets can afford premium subscriptions that offer advanced features, in contrast with smaller institutions with limited funding, which are confined to free versions with fewer capabilities. This financial divide can hinder research output and innovation, as access to robust analytical tools becomes a privilege of wealthier institutions. When added to the existing ethical issues about using LLMs in research (Section~\ref{sec:ethics}), these inequities not only widen the gap between well-funded and less-funded institutions but also deepen the concerns around the use of AI in research, further driving negativity towards their open adoption.

\subsection{Prompt Engineering}\label{sec:prompting}

LLMs can assist us through conversations in natural language, but the output quality highly depends on how we interact with them. In LLM terms, the directive or question posed to the AI tool, designed to elicit a specific response or piece of information, is called a "prompt." Much work and research is done on structuring our prompts for the best results. Many large companies have defined a new job area as a large language model operation, or LLMOps, which includes prompt engineering as a key task. Prompt engineering is currently an art that is very much in rapid development and may become obsolete as LLMs become smarter. While none of the authors are AI researchers or LLM experts, this section summarizes the lessons we learned during our experiments.

The prompt should be clear, concise, and specific to the task. They should be free of ambiguity and provide all the necessary information for the LLM to generate the desired output. For example, "train a logistic regression on the data frame \texttt{df} using 'A,' 'B,' and 'C' as features and y as target" or "plot the correlations between 'x' and 'y' as density contours for every 'region' category." 

Prompts can (and often should) be iterative. If the initial output is unsatisfactory, the user can modify, refine, and follow up on the prompt based on the generated output and prompt the LLM again. This iterative process can help the user guide the LLM towards the desired output and improve the quality of the generated code. Perplexity is currently the only service that seems to automatically request further input from the user, while other services seem to leave this step to the user. 

Sometimes, asking the LLM to generate multiple response versions may be useful. For example, in the previous paragraph, we generated the definition of "prompt" by requesting "Give me three different one-sentence definitions of an LLM prompt.". The final definition is an edited version of one of the responses. 

It is often best to break down complex tasks into smaller, more manageable sub-tasks and prompt the LLM for each sub-task separately. This approach can help avoid overwhelming the LLM and improve the quality of the generated output. One can also ask the LLM to write out the steps of the process and then ask it to elaborate on each step. Finally, you can even request the LLM to ask questions to refine the context if necessary.

When writing a prompt, one should leverage formatting and punctuation to structure the information effectively. Using bullet points or numbered lists can help organize the prompt and make it easier for the LLM to understand the task. One can also enclose a value in quotation marks to indicate that it is a variable name, as we did above. 

When integrated into an IDE (e.g., GitHub Copilot, Codium) or a notebook (e.g., Deepnote), prompts can directly reference existing variables and files. LLMs can reference columns of data frames and their data type to understand better the tasks and produce more accurate results.

For tasks where an "expert" opinion is needed, imbuing a personality to the LLM is useful. Adding a description of the personality to the prompt will change the response. This can be done in various ways, starting with, e.g.,  "act as a copy editor ..." or "act as a scientifically literate critic but not an astronomer." \citet{Mollick2024} discussed that one can, in principle, create complex personalities and demonstrated using them to get feedback on book chapters. In the same way, a user can ask for a particular tone, e.g., "formal," "friendly," "conversational," etc. There is some evidence \citep{politeness} that politeness produces better responses than rudeness, but overly polite language by itself does not guarantee an optimal response. 

It is essential to note that the output quality is susceptible to the prompt. To formulate an effective prompt, the user must understand the task and the data well. Writing an effective prompt could represent a significant inequity in academia, as researchers may not have the expertise to prompt the LLM effectively for a particular task or be fluent in English. Some LLMs can parse prompts in other languages, but we have not tested this feature. Engaging with LLMs in various tasks, "inviting them to every conversation," teaches users how to craft more effective prompts over time. 

Prompt auto-tuning is a feature that could help users optimize their prompts \citep[e.g.][]{Levi2024}. It may solve this challenge, but it is not mature yet and is unavailable in most freely accessible LLMs.

\subsection{LLMs as the Best Available Human}\label{sec:besthuman}

In the course of this experiment, we came across another useful application of LLMs within the context of research that does not quite fit the previous categories, namely using LLMs as the "best available human" \citep{Klein2024}. Generating ideas is a creative process, and within that context, LLMs can serve as invaluable resources as discussion partners for researchers, an advanced version of a rubber duck \citep{Hunt1999} or sounding board \citep{Chen2023}, providing a platform for researchers to articulate their thoughts and ideas. LLMs can act as an always-available colleague to discuss ideas with. Throughout this process, LLMs can assist us in fostering creativity and idea generation. By engaging in a dialogue with an LLM, researchers can gain new perspectives and insights, helping them diversify or refine their thinking and approach to a problem. While LLMs may not possess the same deep expertise as human researchers, they can still offer novel ideas or perspectives, making them a tool for brainstorming and problem-solving. AI-driven tools can facilitate more structured and efficient brainstorming through constant suggestions and prompts. This structured approach can be particularly beneficial in academic research, where generating and refining hypotheses is crucial.

LLMs can generate convincing responses, but without guarantees of factual accuracy -- much like our human colleagues. However, as emphasized earlier, this is where our expertise becomes crucial. Researchers must critically evaluate these responses against reliable sources and empirical data. Scientific knowledge depends on rational reasoning and logical inferences, which require deep, domain-specific expertise. LLMs, lacking the ability to make nuanced value judgments and relational connections, are not (yet) capable of formulating meaningful research questions, designing sound experiments, or fully understanding and articulating their limitations.

\section{Recommendations}\label{sec:recommendations}

It is clear, based on our conversations with colleagues as well as research \citep{astarita2024, geng2024} that LLMs are widely used by many researchers in the field and the Pandora box of generative AI is unlikely to close any time soon. At the same time, the negative attitudes towards such tools make citations unlikely, and the ethics of using these tools for research are frequently murky without clear guidelines from journals, observatories, granting bodies, and professional organizations. This present section attempts to list recommendations astronomers should adhere to maximize the benefits of LLMs while mitigating their limitations \citep[e.g.,][]{Birhane2023,fecher2023}. We divide the recommendations into three sections: for individual researchers, for publishers, and for the community as a whole. We realize that in the fast-moving field of AI, these recommendations may become obsolete very quickly, but we hope they may be nonetheless useful.

\subsection*{Recommendations for individual researchers}

\begin{itemize}
    \item All authors are responsible for the content of their work, whether or not using LLMs or AI tools. Scientists must be aware of this responsibility at all stages of the research process and in all academic tasks.
    \item Researchers must maintain a critical mindset when using AI tools. All outputs of LLMs must be fact-checked and verified against reliable sources, empirical data, or via careful examination. One should use these tools as supplementary aids rather than definitive sources.
    \item Researchers must continuously monitor and evaluate the performance and biases of AI tools and avoid using them for resource allocation where such biases can influence the outputs.
    \item Due to the rapid evolution of AI tools, users must keep up to date on the strengths and weaknesses of LLMs to make informed decisions. 
    \item Researchers should be engaged in creating ethical principles and guidelines for the responsible use of AI in scientific research. 
\end{itemize}

\subsection*{Recommendations for journals}

\begin{itemize}
    \item Journals should remind authors that they bear full responsibility for the content of submitted manuscripts, which must consist of original work produced by the authors. 
    \item Conversely, journals should remind referees they bear full responsibility for the contents of their reviews of the manuscripts of their peers. Furthermore, manuscript authors hold the copyrights to their texts, and it should not be permissible for third parties to upload them to LLMs. 
    \item Journals should implement systems to enforce existing ethical standards, preventing the submission of spam articles or those with fake citations, to protect the publication system from abuse by bad actors. Consequences should exist for bad actors. 
    \item Given the complexities of citing LLMs, journals should evaluate methods for tracking LLM use in manuscripts and reviews. Alternatives to citing LLMs as software may include blinding this information during peer review or collecting it solely for statistical purposes.
    \item Journals must create opportunities for debate within the community and conversations with other disciplines so astronomers can reach a consensus on the rules and guidelines for using LLMs in research and preparing astronomical manuscripts.
\end{itemize}

The same recommendations can apply to time allocation committees (TAC) proposal reviews and similar processes associated with resource allocation.

\subsection*{Recommendation for the astronomical community and professional societies:}

\begin{itemize}
    \item The astronomical community should stay informed about advancements in generative AI and regularly assess their impact on the field.
    \item Professional societies should collect anonymous information on the use of AI in research to assess the impact of these technologies on the scientific process. 
    \item The community should carefully use AI tools as aids and refrain from delegating essential tasks to the scientific community, particularly reviews of all kinds and resource allocation decisions.
    \item The astronomical community and professional societies must create platforms for open discussions and debates, enabling researchers to collaboratively establish guidelines and best practices for using generative AI in research.
    \item The astronomical community should actively engage with other scientific disciplines in discussions about the use of AI and LLMs, ensuring astronomers contribute to broader debates on ethical standards, fairness, and bias in scientific research. Ongoing participation in interdisciplinary dialogue is essential to promote consistent ethical practices and maintain the integrity of scientific work in astronomy.
\end{itemize}

\section{Summary}\label{sec:summary}

Large Language Models represent a transformative technology with the potential to revolutionize how we conduct astronomical research. By leveraging their capabilities in idea generation, literature review, and paper drafting, LLMs can augment the productivity and creativity of astronomers. However, it is crucial to recognize their limitations, such as potential inaccuracies and lack of contextual understanding.

We must use LLMs responsibly and judiciously, combined with human expertise and critical thinking, to maximize their benefits while mitigating risks. The effective use of AI tools requires a combination of technical proficiency and critical thinking skills to interpret and validate the results generated by the AI. Scientists must be able to identify errors, biases, and limitations in the AI-generated outputs and make informed decisions based on their domain knowledge and expertise. As astronomy research continues to evolve, integrating LLMs and their responsible application may be pivotal in driving scientific discoveries and advancing our understanding of the cosmos. 

The goal of this work was to investigate the practical applications of LLMs to research tasks. The work was carried out by a group of astronomers at MPIA with a range of seniority levels and scientific expertise in the Spring of 2024. A wide variety of LLM services were used to perform various tasks as part of our daily activities, as well as pointed experiments. A survey was also circulated among staff at the institute. Below is a summary of our analysis. 

{\bf The potential applications and limitations of using LLMs for astronomy research} and daily tasks are considered in detail in Section \ref{sec:use-cases}. The transformative power of LLMs lies in their ability to streamline various aspects of the research process. LLMs can significantly accelerate literature reviews by summarizing key points, comparing results across publications, and even breaking down disciplinary barriers (Section \ref{subsec:summarizing}). In the realm of coding, LLMs act as intelligent assistants, offering code suggestions, explanations, and debugging support, thereby enhancing productivity and code quality (Section \ref{subsec:code}). Integrating AI tools into data analysis platforms further empowers scientists by automating routine tasks, providing real-time guidance, and facilitating knowledge sharing (Section \ref{subsec:data_analysis}). Moreover, LLMs can contribute to effective science communication by generating visually appealing and informative images for outreach purposes (Section \ref{subsec:outreach}). Across all these tasks, when used responsibly, generative AI platforms are not replacing scientists but allowing them to do their work more effectively.

{\bf However, it is crucial to acknowledge the limitations of LLMs} (Section \ref{sec:limitations}). They generate predictions of the "statistically likely continuations of word sequences (tokens)" \citep{Shanahan2022} based on brute-force iterative training on massive corpora of digital text data, including non-scientific documents and varied code sources. They lack the relational knowledge and nuanced value judgments that are a prerequisite of scientific interpretation. They may occasionally produce inaccurate or incomplete responses, especially when dealing with niche topics or complex concepts. The reliance on LLMs for tasks like summarizing scientific literature can lead to oversimplification and misinterpretation of research findings. In the context of coding, LLMs might suggest flawed or outdated code snippets, necessitating careful review by experienced programmers. Using LLMs for image generation also requires vigilance to ensure scientific accuracy and avoid the dissemination of misleading visuals. In conclusion, LLMs hold immense promise for revolutionizing astronomy research and daily tasks. Still, their responsible and informed use, coupled with human oversight, is paramount to harnessing their full potential while mitigating their limitations. 

{\bf We further discuss the ethical and legal concerns surrounding using LLMs, both in general and specifically within the field of astronomy} in Section \ref{sec:ethics}. The general concerns include copyright infringement in training data, ownership of intellectual property of text generated by LLMs, exploitative labor practices in model training, inherent biases in training data, and the environmental impact of LLMs. We also explore research-specific concerns, such as the impact of LLMs on the scientific process, the integrity of publications, the challenges of enforcing guidelines around LLM use, and the potential for misuse in tasks like proposal and paper reviews. We conclude by highlighting the potential for LLMs to create inequalities in academia due to the cost associated with premium access. The ethical and legal challenges LLMs pose extend far beyond astronomy, reflecting broader societal debates across academia, education, legal systems, and philosophy as we collectively navigate the impact of these technologies. The astronomical community must collaborate with other fields to address these complex challenges and ensure that innovation is balanced with ethical responsibility, integrity, and fairness.

{\bf To effectively utilize LLMs, users need a combination of technical knowledge and critical thinking skills.} It's crucial to understand that LLMs generate predictions based on statistical patterns rather than true comprehension, meaning they may produce inaccurate or incomplete responses, particularly in specialized or complex areas. Users must be able to critically assess these outputs against reliable sources and empirical data. Additionally, domain expertise is needed to verify the relevance and accuracy of the results. Crafting well-structured prompts is also essential to maximize the quality of responses. Ethical awareness, including the potential for bias in training data and the implications of intellectual property, is also important. Beyond their individual work, researchers, especially those with a deep understanding of generative AI systems, must be proactive community members and participate in creating standards that mitigate the risks associated with LLMs.

{\bf Finally, we conclude the paper with an attempt at recommendations} for individual researchers, publishers and the astronomical community as a whole (Section \ref{sec:recommendations}). In these recommendations we try to strike a balance between the reality that LLMs are likely widely-used, the psychology of admitting to having used LLMs and the ethics of responsible research. However, generative AI is a fast-moving field and we expect a significant evolution in this area, which may make these recommendation moot in the short term. 

In conclusion, the scientific community should adopt a responsible and informed approach to using LLMs in research. Scientists must view LLMs, and more broadly generative AI technologies, as exploratory tools. The scientific community must continue to enforce the rigor of scientific works.

\section*{Acknowledgements}
\input{acknowledgements}

\section*{Data Availability}


For this work, we relied on external public services and versions listed in Table\,\ref{tab:services_used}. In addition, we collected responses from a survey provided in Appendix \ref{sec:survey_questions}.



\bibliographystyle{rasti}
\bibliography{main}



\appendix

\section{LLM services used in this work}\label{sec:services}

Table \ref{tab:services_used} briefly lists the services we used in this work. This section details their characteristics.

\begin{itemize}
    \item{{\bf ChatGPT}\footnote{\url{https://chatgpt.com}} is a language model by OpenAI, featuring a context window of 4,096 tokens/175 billion parameters (ChatGPT-3.5) or 8,192 tokens/$>1$ trillion parameters (ChatGPT-4 and 4.5, though token windows of ~128,000 are available in newer models), designed for generating coherent, context-aware text across a variety of applications. We tested both the 3.5 and 4. The 4.5 version of the model was not available during the time frame of the experiments. One of the main limitations of an earlier version of the model was the lack of internet access and the training data's cut-off date. Newer versions (v4.0 and above) can access the internet.}
    \item{{\bf Gemini}\footnote{\url{gemini.google.com}}  (formerly Bard) model developed by Google AI, which comes with a standard context window of 128,000 tokens in its standard tier and up to a million tokens for the advanced tier.}
    \item{{\bf Claude2}\footnote{\url{https://claude.ai}} from Anthropic has 137 billion parameters. Its strengths are in improved factual accuracy and reasoning compared to Claude v1, longer context window (100 000 tokens, $\sim$ the length of a book), weaknesses in limited access, and is in beta testing, limited beta access.}
    \item{{\bf Perplexity}\footnote{\url{https://www.perplexity.ai}} is a conversational LLM-powered search engine that responds with a combination of generated text and internet resources. The free version relies on the company's own LLM, while the paid version offers access to GPT-4, Claude 3.5, Mistral Large, Llama 3, and an Experimental Perplexity Model. The context window ranges from 4,000 to 32,000 tokens, and models range from 8 billion to 70 billion parameters. The benefit of Perplexity is that it is integrated with internet access and can provide citations and links to further resources on the internet.}
    \item{{\bf You.com}\footnote{\url{https://you.com}} is also an LLM-powered search engine with a focus on personalization and access to the internet and various AI-powered tools. At the time of writing, the free interface offers unlimited access to their custom LLM engine and daily limited access to different leading AI models, such as GPT-4 and Turbo, Claude Instant, 2, 3 Opus and 3 Sonnet, Gemini Pro, Zephyr, and others. The interface offers different modes, including Smart (free), Research, Genius, and Creative, which provide specializations. For example, the Smart mode ``delivers quick, reliable responses with live web access, citations, and sources.'' Their paid version offers unlimited access to all models and tools, including the ability to train custom models.}
    \item{{\bf Copilot}\footnote{\url{https://github.com/features/copilot}} is an LLM-powered code completion tool developed by GitHub and OpenAI based on a 1.75-billion parameter model. The LLM model used data from public GitHub repositories to train, which provided diverse implementations of similar problems and a broad range of programming languages. The tool is available by subscription and free for students and educators. Copilot works as an extension in integrated development environments (IDE) and works best for Python, JavaScript, TypeScript, Ruby, and Go languages. Copilot can convert instructions from natural language into code, create documentation and test, and explain code, among other tasks. The context of the model can include either the file open in the IDE or (more recently) the full project.}
    \item{{\bf Codeium}\footnote{\url{https://codeium.com/}} is a free (for individual users) alternative to GitHub Copilot that provides similar functionalities. Codeium uses its own proprietary language model (LLM) to generate code snippets and suggestions. The paid service (for teams) provides access to GPT4 and a better context awareness window.}
    \item{{\bf Grammarly}\footnote{\url{https://app.grammarly.com/}} is an AI-powered writing assistant that helps users improve their writing by providing real-time suggestions on grammar, spelling, punctuation, and style. It integrates with various platforms, including web browsers, word processors, and email clients, making it convenient to use across different writing tasks. Additionally, Grammarly offers advanced features such as tone detection and plagiarism checking. LLM-based tools include paraphrasing in different styles, generating outlines, summarizing, and generating text based on a prompt. Detailed information about the underlying model is not publicly available. }
    \item{{\bf DeepL}\footnote{\url{https://www.deepl.com/}} is a neural network-based machine translation service which currently supports 33 languages. Since 2023, the company has also offered DeepL Write, a monolingual tool to improve texts in English and German. The tool is an editor that edits already written text, corrects typos and grammatical errors, and suggests better wording. A free version is limited to a certain number of words. The interface also offers different writing styles. DeepL aims to distinguish itself with the targetted acquisition of specific datasets and a much smaller model tailored for the specific tasks it supports. Detailed information about the underlying model is not publicly available.}
    \item{{\bf Midjourney}\footnote{\url{https://www.midjourney.com/}} excels at creating images based on text prompts, focusing on artistic styles and transformations. It can also apply different artistic styles to existing images, allowing users to experiment with various aesthetics.}
    \item{{\bf DALL-E}\footnote{\url{https://www.openai.com/dall-e}} is developed by OpenAI and generates images directly from textual descriptions using a generative pretrained transformer model. The latest version, DALL-E 3, has improved image quality and adherence to prompts.}
\end{itemize}

\section{Survey Questions and Responses}\label{sec:survey_questions}

Questions marked with an asterisk are required. Answers marked with a square are multiple-choice, while answers marked with a circle allow only one selection.

{\bf General questions}

\begin{itemize}

\item What is your academic level/position? * (27 responses)
    \begin{itemize}
        \itemr Professor \dotfill 7\%
        \itemr Staff member \dotfill 30\%
        \itemr PostDoc \dotfill 11\%
        \itemr PhD student \dotfill 26\%
        \itemr Masters student \dotfill 11\%
        \itemr Bachelor student \dotfill 11\%
        \itemr Intern \dotfill 4\%
        \itemr Other: \dotfill 0\%
    \end{itemize}

\item Which LLMs have you used? * (27 responses)
    \begin{itemize}
        \itemc ChatGPT 3.5 \dotfill 89\%
        \itemc ChatGPT 4 \dotfill 44\%
        \itemc Bard/Gemini1.0 \dotfill 7\% 
        \itemc Grammarly \dotfill 41\%
        \itemc Perplexity \dotfill 7\%
        \itemc Github Copilot \dotfill 37\% 
        \itemc None \dotfill 4\%
        \itemc Other: \dotfill DeepL (4\%), Codeium (4\%)
    \end{itemize}

\item How frequently do you use LLMs in your work? * (27 responses)
    \begin{itemize}
        \itemr Daily \dotfill 30\%
        \itemr Several times in a week \dotfill 44\%
        \itemr Monthly \dotfill 7\%
        \itemr Rarely (a few times a year) \dotfill 15\%
        \itemr Never \dotfill 4\%
    \end{itemize}

\item For what do you use LLMs? (25 responses)
    \begin{itemize}
        \itemc Coding Assistance \dotfill 92\%
        \itemc Writing Assistance (proposals, etc.) \dotfill 72\%
        \itemc Answering (scientific) questions \dotfill 28\%
    \end{itemize}

\item Do you find it difficult to use LLMs? * (27 responses)
    \begin{itemize}
        \itemr Very difficult \dotfill 0\%
        \itemr Difficult \dotfill 0\%
        \itemr Neither difficult nor easy \dotfill 30\%
        \itemr Easy \dotfill 44\%
        \itemr Very easy \dotfill 26\% 
    \end{itemize}

\item What is difficult/keeps you from using LLMs? (8 responses)
    \begin{itemize}
        \itemc Ethics \dotfill 50\% (4 responses)
        \itemc Learning curve \dotfill 38\% (3 responses)
        \itemc Accessibility \dotfill 50\% (4 responses)
    \end{itemize}

\end{itemize}

{\bf Coding}: This section is meant to give a better insight in how the LLMs are used in practice and to find out how they improve your ability to code for work.

\begin{itemize}

\item Do you use LLMs for coding assistance? * (27 responses)
    \begin{itemize}
        \itemr Yes \dotfill 89\%
        \itemr No \dotfill 11\%
    \end{itemize}

\item How accurate were your coding prompt results? (25 responses)
    \begin{itemize}
        \itemr 100 \% \dotfill 8\%
        \itemr 80-99 \% \dotfill 20\%
        \itemr 60-80 \% \dotfill 44\%
        \itemr 40-60 \% \dotfill 20\%
        \itemr 20-40 \% \dotfill 0\%
        \itemr 1-20 \% \dotfill  4\%
        \itemr 0 \% \dotfill 4\%
    \end{itemize}

\item Using an LLM helps me… (Mark any that apply to you) (24 responses)
    \begin{itemize}
        \itemc …write more code in less time \dotfill 63\%
        \itemc …complete coding tasks faster \dotfill 83\%
        \itemc …identify and fix bugs more efficiently \dotfill 58\%
        \itemc …produce higher quality code (e.g., fewer errors, better readability, faster) \dotfill 42\%
        \itemc …write code that compiles and runs without errors more often \dotfill 8\%
        \itemc …improve my coding knowledge \dotfill 54\%
        \itemc …make difficult coding tasks easier \dotfill 29\%
        \itemc …increase my confidence in the correctness and efficiency of the code I write \dotfill 13\%
        \itemc …learn new coding concepts or techniques \dotfill 54\%
        \itemc …in inspiring me to try new approaches to coding problems \dotfill 21\%
        \itemc …write the code for a theoretical concept \dotfill 13\%
        \itemc …generate code for doing advanced tasks (such as data cleaning) \dotfill 21\%
    \end{itemize}

\item How much of your code is assisted/inspired by LLMs? (26 responses)
    \begin{itemize}
        \itemr 100 \% \dotfill 0\%
        \itemr 80-99 \% \dotfill 8\%
        \itemr 60-80 \% \dotfill 8\%
        \itemr 40-60 \% \dotfill 15\%
        \itemr 20-40 \% \dotfill 20\%
        \itemr 1-20 \% \dotfill 42\%
        \itemr 0 \% \dotfill 8\%
    \end{itemize}

\item How many times has an LLM suggested an advanced snippet (e.g., decorator in Python)? (24 responses)
    \begin{itemize}
        \itemr Often \dotfill 4\%
        \itemr Sometimes \dotfill 25\% 
        \itemr Rarely \dotfill 33\%
        \itemr Never \dotfill 38\%
    \end{itemize}

\item Have you used LLMs to convert code from one programming language to another? (23 responses)
    \begin{itemize}
        \itemr Yes \dotfill 17\%
        \itemr No \dotfill 83\%
    \end{itemize}

\item What programming language did you convert code from? (5 responses)
    \begin{itemize}
        \itemc Python \dotfill 20\% (1 response)
        \itemc C++ \dotfill 0\%
        \itemc C \dotfill  0\%
        \itemc Julia \dotfill 20\% (1 response)
        \itemc IDL \dotfill 40\% (2 responses)
        \itemc Fortran \dotfill 20\% (1 response)
        \itemc Rust \dotfill  0\%
        \itemc Other: \dotfill  0\%
    \end{itemize}

\item What programming language did you convert code to? (4 responses)
    \begin{itemize}
        \itemc Python \dotfill 100\% (4 responses)
        \itemc C++ \dotfill  0\%
        \itemc C \dotfill  0\%
        \itemc Julia \dotfill  0\%
        \itemc IDL \dotfill  0\%
        \itemc Fortran \dotfill  0\%
        \itemc Rust \dotfill  0\%
        \itemc Other: \dotfill  0\%
    \end{itemize}

\item How accurate was such a conversion? (4 responses)
    \begin{itemize}
        \itemr 100 \% \dotfill 25\%  (1 response)
        \itemr 80-99 \% \dotfill 0\%
        \itemr 60-80 \% \dotfill  0\%
        \itemr 40-60 \% \dotfill  50\%  (2 responses)
        \itemr 20-40 \% \dotfill  25\%  (1 response)
        \itemr 1-20 \% \dotfill  0\%
        \itemr 0 \% \dotfill  0\%
    \end{itemize}

\end{itemize}

{\bf Text writing: }  This section is meant to give a better insight into how the LLMs are used in practice and to find out how they improve your ability to write scientific texts (e.g., proposals, papers, applications, etc.)

\begin{itemize}

\item Do you use LLMs for writing or improving academic texts such as papers/application essays/proposals/etc? * (27 responses)
    \begin{itemize}
        \itemr Yes \dotfill 59\%
        \itemr No \dotfill 41\%
    \end{itemize}

\item How much of your writing is assisted/inspired by LLMs? (22 responses)
    \begin{itemize}
        \itemr 100 \% \dotfill 0\%
        \itemr 80-99 \% \dotfill 0\%
        \itemr 60-80 \% \dotfill 0\%
        \itemr 40-60 \% \dotfill 14\%
        \itemr 20-40 \% \dotfill 18\%
        \itemr 1-20 \% \dotfill 46\%
        \itemr 0 \% \dotfill 23\%
    \end{itemize}

\item How accurate were your text prompt results? (20 responses)
    \begin{itemize}
        \itemr 100 \% \dotfill 0\%
        \itemr 80-99 \% \dotfill 15\%
        \itemr 60-80 \% \dotfill 25\%
        \itemr 40-60 \% \dotfill 20\%
        \itemr 20-40 \% \dotfill 30\%
        \itemr 1-20 \% \dotfill 0\%
        \itemr 0 \% \dotfill 10\%
    \end{itemize}

\item Have you used LLMs for peer review or manuscript editing? (21 responses)
    \begin{itemize}
        \itemr Yes \dotfill 33\%
        \itemr No \dotfill 67\%
    \end{itemize}

\end{itemize}

{\bf Problem solving}: This is a short section on the general use of solving either homework/work-related problems (any problem that is not coding or text generation).

\begin{itemize}

\item How often would you say an LLM can help you solve a scientific problem? * (27 responses)
    \begin{itemize}
        \itemr Often \dotfill 7\%
        \itemr Sometimes \dotfill 33\% 
        \itemr Rarely \dotfill 30\%
        \itemr Never \dotfill 30\%
    \end{itemize}

\item How helpful is an LLM for solving problems? (23 responses)
    \begin{itemize}
        \itemr Very helpful \dotfill 9\%
        \itemr Helpful \dotfill 57\%
        \itemr Not very helpful \dotfill 30\%
        \itemr No help at all \dotfill 4\%
    \end{itemize}

\end{itemize}

{\bf Outreach/Visualisation}: This section is mainly on outreach but contains some image/plot generation questions regarding LLMs.

\begin{itemize}

\item Have you used LLMs for Outreach? * (27 responses)
    \begin{itemize}
        \itemr Yes \dotfill 7\%
        \itemr No \dotfill 93\%
    \end{itemize}

\item You used LLMs for… (Mark any that apply to you) (9 responses)
    \begin{itemize}
        \itemc Generating summaries \dotfill 100\% (9 responses)
        \itemc Alt-text or captions \dotfill 22\% (2 responses)
        \itemc Images \dotfill 22\% (2 responses)
        \itemc Titles or acronyms \dotfill 33\% (3 responses)
        \itemc Other: \dotfill 11\% (translation, 1 response)
    \end{itemize}

\item Have you used LLMs to generate educational images? (20 responses)
    \begin{itemize}
        \itemr Yes \dotfill 15\%
        \itemr No \dotfill 85\%
    \end{itemize}

\item Have you used LLMs to generate scientific plots? (20 responses)
    \begin{itemize}
        \itemr Yes \dotfill 25\%
        \itemr No \dotfill 75\%
    \end{itemize}

\item How many prompts did you need until such an image/plot was satisfactory? (6 responses)
    \begin{itemize}
        \itemr 1 \dotfill 0\%
        \itemr 2-4 \dotfill 33\% (2 responses)
        \itemr 4-6 \dotfill 17\% (1 response)
        \itemr 6-8 \dotfill 0\%
        \itemr More than 8 \dotfill 50\% (3 responses)
    \end{itemize}

\item Was the generated image scientifically accurate? (5 responses)
    \begin{itemize}
        \itemr Yes \dotfill 40\%
        \itemr No \dotfill 60\%
    \end{itemize}

\item How accurate were your outreach prompting results? (3 responses)
    \begin{itemize}
        \itemr 100 \% \dotfill 0\%
        \itemr 80-99 \% \dotfill 0\%
        \itemr 60-80 \% \dotfill 33\% (1 response)
        \itemr 40-60 \% \dotfill 33\% (1 response)
        \itemr 20-40 \% \dotfill 33\% (1 response) 
        \itemr 1-20 \% \dotfill 0\%
        \itemr 0 \% \dotfill 0\%
    \end{itemize}

\end{itemize}

{\bf Ethics}: What are the ethical implications of using LLMs?

\begin{itemize}

\item Do you think it is needed to credit LLMs in papers? * (27 responses)
    \begin{itemize}
        \itemr Yes \dotfill 78\%
        \itemr No \dotfill 22\%
    \end{itemize}

\item How often do you credit LLMs in papers? (18 responses)
    \begin{itemize}
        \itemr Every time I use them \dotfill 28\%
        \itemr Sometimes \dotfill 11\%
        \itemr Rarely \dotfill 61\%
        \itemr Never 0\%
    \end{itemize}

\item How do you address potential biases or inaccuracies in information provided by LLMs? (25 responses)
    \begin{itemize}
        \itemc Double check information provided by an LLM \dotfill 84\%
        \itemc Change the prompt to the LLM to get different results and compare them \dotfill 40\%
        \itemc Use a different LLM \dotfill 24\%
        \itemc Other:
        \begin{itemize}
            \item Think for myself \dotfill 4\%
            \item Use LLMs for inspiration and re-write text \dotfill 4\%
            \item Only use LLMs in situations where output can be checked with personal knowledge \dotfill 4\%
        \end{itemize}
    \end{itemize}

\end{itemize}

{\bf Overall Satisfaction}: A few final questions on the LLMs.

\begin{itemize}

\item Overall, how satisfied are you with the current capabilities of LLMs in your academic work? (27 responses)
    \begin{itemize}
        \itemr Very Satisfied \dotfill 7\%
        \itemr Satisfied \dotfill 41\%
        \itemr Neutral \dotfill 41\%
        \itemr Not Satisfied \dotfill 7\% 
        \itemr Not at all Satisfied \dotfill 4\%
    \end{itemize}

\item How satisfied are you with the LLM as a tool for coding? (25 responses)
    \begin{itemize}
        \itemr Very Satisfied \dotfill 32\%
        \itemr Satisfied \dotfill 44\%
        \itemr Neutral \dotfill 12\%
        \itemr Not Satisfied \dotfill 8\% 
        \itemr Not at all Satisfied \dotfill 4\%
    \end{itemize}

\item How satisfied are you with the LLM as a tool for text writing or problem-solving? (24 responses)
    \begin{itemize}
        \itemr Very Satisfied \dotfill 8\%
        \itemr Satisfied \dotfill 50\%
        \itemr Neutral \dotfill 25\%
        \itemr Not Satisfied \dotfill 8\%
        \itemr Not at all Satisfied \dotfill 8\% 
    \end{itemize}

\item Is there a need to buy the pro version of such tools? * (27 responses)
    \begin{itemize}
        \itemr Yes \dotfill 59\%
        \itemr No \dotfill 41\%
    \end{itemize}

\item Which tool would you buy the pro version of? (21 responses)
    \begin{itemize}
        \itemc ChatGPT \dotfill 72\%
        \itemc Grammarly \dotfill 24\%
        \itemc Bard/Gemini \dotfill 5\%
        \itemc Perplexity \dotfill 5\%
        \itemc Github Copilot \dotfill 57\%
        \itemc Other: \dotfill Writeful (1\%), Not sure (1\%)
    \end{itemize}

\item How do you see the role of LLMs evolving in academia in the future? (24 responses)
    \begin{itemize}
        \itemc They will be an integral part of science \dotfill 83\%
        \itemc They will be a bad influence on science overall \dotfill 25\% 
        \itemc They will have a good influence on science overall \dotfill 42\%
    \end{itemize}

\item What improvements or features would you like to see in future versions of LLMs? (8 responses)
    \begin{itemize}
        \item Some means to determine the accuracy of results. 
        
        \item Information about the sources that went into the models.
        
        \item Some kind of indicator about the "confidence" of the LLM's answers. Sometimes it clearly can change the answer from one prompt to another. But it definitely will defend its position/mention potential mistakes when it is super "confident" in its answers.
        
        \item Less hallucination, more agency, more logical reasoning.
        
        \item Eliminate the training time lag of several years.
        
        \item Better and more flexible capabilities in parsing and suggestions when it comes to bigger projects (I know that copilot looks at the last 5 files, but that's sometimes not sufficient).
        
        \item I've answered that LLM will have both bad and good influence on science overall but not because of LLM, it's just a tool, but I think it's important to teach people about it's limitations and caveats. I find LLM very helpful but I also notice where it does not provide correct information and can be misleading, I notice this cause I've acquired enough expertise before LLM became widespread and I am not sure how it will impact the studying process for future generations without proper education on how it works. Overall, I am quite positive about it, I don't think we should demonize it or stop young people from using but they should be aware enough of possible biases. I'm not super sure which improvements to expect; it's not magic, and I don't think it will ever be with no limitations.
        
        \item When they honestly answer that they can't answer scientific questions instead of lying about "facts."

        \item Referencing the sources.
    \end{itemize}

\end{itemize}

\section{Examples of prompts and answers from LLM services}

This section presents a few examples of prompts and answers we collected from LLM services that we refer to in the main text.

\begin{table*}
\caption{Image generation process for a globular cluster with tidal tails using ChatGPT}
\label{tab:outreach}
\begin{tabular}{| m{0.18\linewidth} |  m{0.6\linewidth} | m{0.18\linewidth} |}
\hline
Prompt                                                                                                                                                & Response                                                                                                                                                                                                                                                                                                                                                                                                              & Image \\ \hline
Can you draw a scientific image of a tidally disrupted globular cluster with tidal tails?                                                             & I've created a scientifically accurate depiction of a tidally disrupted globular cluster with tidal tails, highlighting the gravitational forces at play and the dynamic, chaotic nature of these interactions against the backdrop of space.                                                                                                                                                                         &  \includegraphics[width=0.18\textwidth]{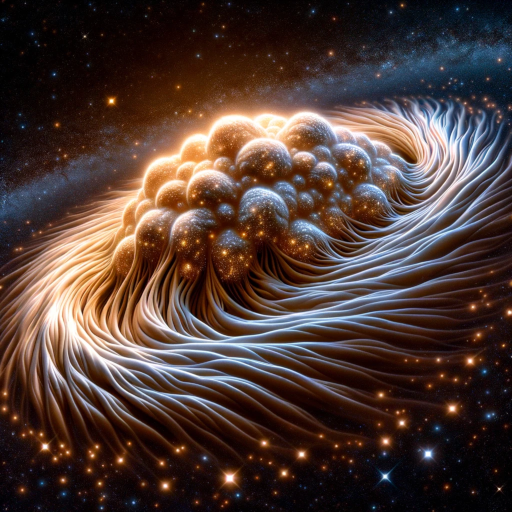}
\\ \hline
That doesn't look quite right to me, can you try again?                                                                                                & I've created a new image with a focus on depicting a tidally disrupted globular cluster and the formation of tidal tails, emphasizing the streams of stars and stellar debris being pulled away from the cluster's core against the backdrop of the cosmos.                                                                                                                                                           & \includegraphics[width=0.18\textwidth]{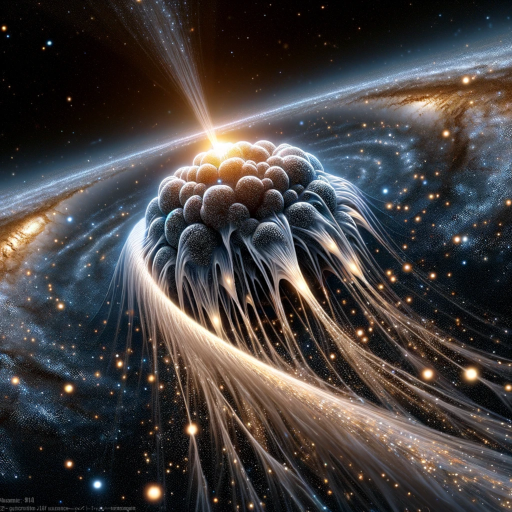}
\\ \hline
Can you draw a globular cluster in the Milky Way galaxy                                                                                               & I've created a scientifically accurate depiction of a globular cluster within the Milky Way Galaxy, emphasizing the dense concentration of stars within the cluster against the panoramic backdrop of the galaxy. This visualization captures the majesty and vast expanse of the Milky Way, highlighting the significance and beauty of the globular cluster within one of the galaxy's spiral arms.                 &      
\includegraphics[width=0.18\textwidth]{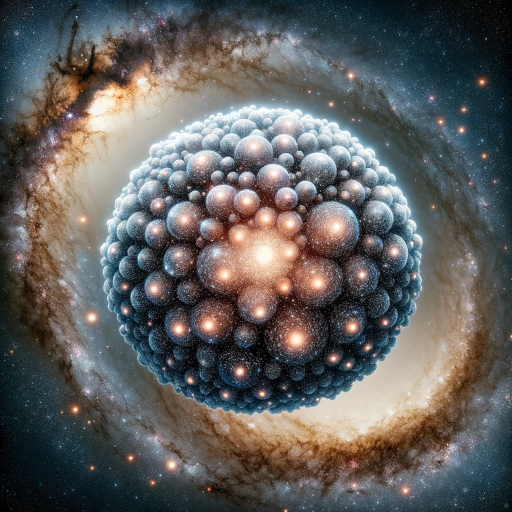}
\\ \hline
That is not what a globular cluster looks like. Please try again.                                                                                     & I've created a new image focusing on a more accurate representation of a globular cluster, depicted as a densely packed, spherical assembly of stars within the Milky Way Galaxy. This visualization showcases the intricate structure and the radiant glow from the numerous stars in the cluster, set against a backdrop that subtly represents the Milky Way, emphasizing the cluster's beauty and cosmic context. &   
\includegraphics[width=0.18\textwidth]{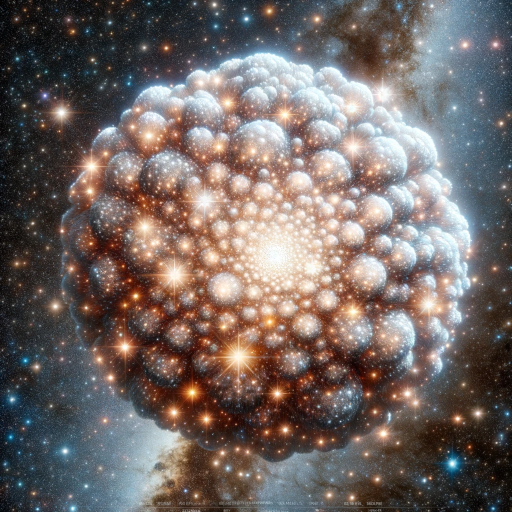}
\\ \hline
Here is an image of a globular cluster with tidal tails. Can you use this to make other images?                                                       & Thank you for providing the example image of a globular cluster with tidal tails. Now that I have a reference, I can use this to create other images. What would you like the next image to feature or emphasize?                                                                                                                                                                                                     &       
\includegraphics[width=0.18\textwidth]{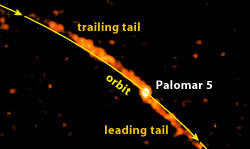} M. Odenkirchen and E. Grebel.
\\ \hline
I'd like a picture of a globular cluster with tidal tails that is suitable for an outreach talk; it should look aesthetically pleasing and impressive & Here is an illustration of a globular cluster with tidal tails, designed to be aesthetically pleasing and impressive for an outreach talk. The image emphasizes the grandeur of the cluster and the elegant flow of the tidal tails, set against a vibrant backdrop of the cosmos.                                                                                                                                    &       
\includegraphics[width=0.18\textwidth]{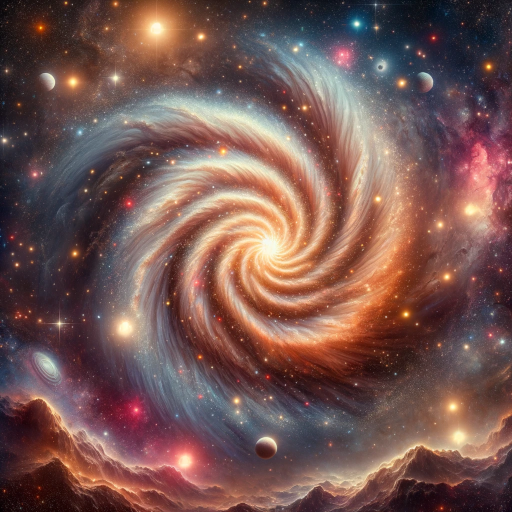}
\\ \hline
That's too artistic, can you make it more like the reference image I sent?                                                                            & I've created an illustration that aligns more closely with the reference image you provided, featuring a globular cluster with tidal tails against a dark space backdrop, designed to be captivating for an outreach talk. The tails are rendered with a visual flair to capture the audience's attention while maintaining a nod to the scientific nature of the subject.                                            &      \includegraphics[width=0.18\textwidth]{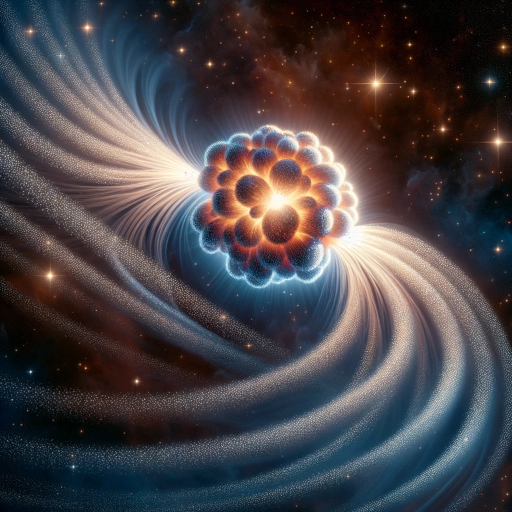}\\ \hline
\end{tabular}
\end{table*}


\bsp	
\label{lastpage}
\end{document}

%% file: authors.tex
\author[M. Fouesneau et al.]
{Morgan Fouesneau\orcit{0000-0001-9256-5516}$^{1}$\thanks{E-mail: fouesneau@mpia.de},
Ivelina G. Momcheva\orcit{0000-0003-1665-2073}$^{1}$, \and 
Urmila Chadayammuri\orcit{0000-0003-2521-506X}$^{1}$, 
Mariia Demianenko\orcit{0000-0002-8297-6386}$^{1,2}$,  
Antoine Dumont\orcit{0000-0003-0234-3376}$^{1}$,
Raphael E. Hviding\orcit{0000-0002-4684-9005}$^{1}$, \and 
K. Angelique Kahle\orcit{0000-0001-7714-7551}$^{1}$, 
Nadiia Pulatova$^{1}$, 
Bhavesh Rajpoot\orcit{0009-0004-9729-6377}$^{1,3}$,
Marten B. Scheuck\orcit{0000-0002-6108-6931}$^{1}$, \and
Rhys Seeburger\orcit{0000-0001-8898-9463}$^{1}$, 
Dmitry Semenov\orcit{0000-0002-3913-7114}$^{1}$, 
Jaime~I. Villase\~{n}or\orcit{0000-0002-7984-1675}$^{1}$
\\
\\
$^{1}$Max-Planck-Institut für Astronomie (MPIA), Königstuhl 17, 69117 Heidelberg, Germany,\\
$^{2}$ Astroinformatics, Heidelberg Institute for Theoretical Studies, Schloss-Wolfsbrunnenweg 35, 69118 Heidelberg, Germany\\
$^{3}$ Heidelberg University, Grabengasse 1, 69117 Heidelberg, Germany
}

%% file: acknowledgements.tex
%
%
This work used various AI tools, in particular, Perplexity AI (\url{https://www.perplexity.ai/}), Open AI GPT-4 (\url{https://openai.com/research/gpt-4}), Google AI Gemini (\url{https://gemini.google.com/}), search engine You.com (\url{https://you.com}), 
GitHub Copilot (\url{https://github.com/features/copilot}), Grammarly (https://grammarly.com), DeepL (\url{https://www.deepl.com/}), and Codeium (\url{https://codeium.com/})
This publication made extensive use of the online authoring Overleaf platform (\url{https://www.overleaf.com/}).\\
This research was supported by the Max Plank Institute for Astronomy.